\documentclass[]{tWRM2e_arxiv}
\usepackage{amssymb,amsbsy,amsmath,amsfonts,amssymb,amscd}
\usepackage{latexsym,euscript,exscale,epic,eepic,epsfig}
\usepackage{color}

\renewcommand{\atop}[2]{\genfrac{}{}{0pt}{}{#1}{#2}}

\begin{document}

\doi{10.1080/1745503YYxxxxxxxx}
\issn{1745-5049}
\issnp{0278-1077}
\jvol{00} \jnum{00} \jyear{2011} \jmonth{00}

\markboth{Y.-P. Pellegrini, P. Thibaudeau and B. Stout}{Momentum-dependent electromagnetic T-matrix}

\title{\itshape The Off-Shell Electromagnetic T-matrix:\\ momentum-dependent scattering from spherical inclusions with both dielectric and magnetic contrast}

\author{
Yves-Patrick Pellegrini$^{\rm a}$$^{\ast}$
\thanks{$^\ast$Corresponding author. Email: yves-patrick.pellegrini@cea.fr\vspace{6pt}}
Pascal Thibaudeau$^{\rm b}$ and Brian Stout$^{\rm c}$\\
\vspace{6pt}
$^{\rm a}${\em{CEA, DAM, DIF, F-91297 Arpajon, France}};\\
$^{\rm b}${\em{CEA, DAM, Le Ripault, BP 16, F-37260 Monts, France}};\\
$^{\rm c}${\em{Institut Fresnel, Universit\'e d'Aix-Marseille, CNRS, F-13397 Marseille, France}}\\
\vspace{6pt}\received{January 12, 2011} }

\maketitle

\begin{abstract}
The momentum- and frequency-dependent T-matrix operator for the scattering of electromagnetic waves by a dielectric/conducting and para- or diamagnetic sphere is derived as a Mie-type series, and presented in a compact form emphasizing various symmetry properties, notably the unitarity identity. This result extends to magnetic properties one previously obtained for purely dielectric contrasts by other authors. Several situations useful to spatially-dispersive effective-medium approximations to one-body order are examined. Partial summation of the Mie series is achieved in the case of elastic scattering.
\bigskip

\bigskip
\end{abstract}

\section{Introduction}
\label{i}
The {\em transition operator}, or $T$-matrix, of a scatterer is a basic building block of time-harmonic theories of single or multiple scattering \cite{VARA80}. It embodies information about the overall polarization-dependent response of a finite object, in terms of the incident and scattered momenta. A recent bibliographical review devoted to the use of T-matrices in electromagnetism \cite{MISC04} (mostly of the ``on-shell'' variety, see definition below) illustrates its key importance in the treatment of the response of heterogeneous media of various natures. Excepting scatterers of the simplest forms however, the T-matrix is generally a painstaking object to compute, and often leads to systems of equations that can be numerically problematic (\textit{e.g.}, Ref.\ \cite{WATE07} and references therein).

In the vast majority of treatments, the T-matrix is computed {\em on-shell}. This terminology, borrowed from particle physics, indicates that the incident and scattered momenta have their norms fixed by the dispersion relation of the host medium. To be precise, the on-shell case corresponds to taking $|\mathbf{k}_1|=|\mathbf{k}_2|=k_m$ with $k_m=(\omega/c)\sqrt{\varepsilon_m\mu_m}$, where $\mathbf{k}_2$ (resp. $\mathbf{k}_1$) stands for the arriving (resp. departing) wavevector, $\omega$ the angular frequency of the incident wave, $c$ the velocity of light in vacuum, and $\varepsilon_m$ and $\mu_m$ are, respectively, the complex relative (frequency-dependent) dielectric permittivity (including conduction) and the para- or diamagnetic permeability in the embedding medium. These latter quantities are assumed homogeneous and isotropic. With on-shell T-matrices, a scattering system can only be examined ``from outside'', \textit{i.e.} by sending and receiving signals from the host medium \cite{MORO97}. However, the modern developments of multiple scattering theory (\textit{e.g.}, \cite{TSAN80,TSAN81}) have made it clear that a \emph{fully general solution} of the scattering problem in the bulk of a heterogeneous system requires that the T-matrix be computed \emph{off shell}, that is, with norms of the arriving and departing momenta, $k_2\equiv|\mathbf{k}_2|$ and $k_1\equiv|\mathbf{k}_1|$ respectively, being arbitrary (not being constrained to equal $k_m$). The full generality provided by the off-shell formalism allows one to study of arbitrary fields in bulk heterogeneous media, and also to deal with interface problems for which approximate techniques are available, see references in \cite{AGRA84}.

Moreover, off-shell computations are the proper context for discussing \emph{spatial non-locality} (also referred to as spatial dispersion) in bulk response of heterogeneous random media \cite{BARR07}. This allows for theoretical investigations of the various propagation modes that can arise as a direct consequence of the finite size of heterogeneities. Information on these modes (most of them strongly attenuated) can be obtained either by direct computation of such ``leaky modes'', or by computing the density of states using the imaginary part of the Green's function \cite{SHEN95}.

Although spatial dispersion in the electrodynamics of crystals is an old sub-domain of solid-state physics \cite{AGRA84,FUCH92}, heterogeneity-induced spatial dispersion in random media is a less understood matter, for which many points remain to be clarified \cite{BARR07}. More importantly, while being for a long time a pure theoretical preoccupation \cite{KARA64,KELL66,BERA70,DIEN76,BART95,BALI95,SHEN95,PELL97b,PELL97c}, it has now acquired some experimental substance in acoustics \cite{LIU90}, electromagnetism in random media \cite{HESP01}, and for electromagnetic metamaterials, \textit{e.g.}, \cite{OBRI02,PROS06}.

Given the complexity of off-shell T-matrices as compared to their on-shell counterparts, we restrict ourselves to the simplest three-dimensional case of a single sphere.  Analytical results for off-shell transition operators in classical physics are the scalar $T$-matrix for acoustic scattering \cite{TSAN81,KIRK85,COND86} and the tensor electromagnetic $T$-matrix for a purely dielectric sphere \cite{TSAN80}. To our knowledge however, an explicit expression similar to that in Ref.\ \cite{TSAN80} for a sphere with both dielectric {\em and} magnetic contrast with respect to its embedding medium has not previously been available in the literature. Since spatial dispersion implies the existence of an effective magnetic-like response even in dielectric media, as has been observed by a number of  authors (\textit{e.g.}, \cite{LAND84,BARR07}), the dielectric \emph{and} magnetic case is a necessary milestone on the road towards a realistic frequency-dependent self-consistent effective-medium theory. Indeed, despite the considerable amount of work having addressed this issue (see Refs.\ \cite{BARR07,SHEN95} and references therein) the latter question remains unsettled.

This paper is devoted to presenting the expression of the off-shell T-matrix operator of a sphere with arbitrary dielectric and magnetic contrast, in the form of a Mie series expansion. This result was derived more than a decade ago \cite{THIB97} but remained unpublished, although having been announced in Refs.\ \cite{PELL97b,PELL97a}. It should be mentioned that by about the same time, Tip independently considered this same off-shell case in an abstract mathematical framework \cite{TIP97}, but gave explicit results for a vacuum background only, and it is unclear how his result compares to ours. In the following, we exclude the situation of a nonzero applied constant magnetic field. An extension of the off-shell T-matrix formalism in this case has recently been put forward and exploited in connection with electromagnetic wave propagation in magnetochiral media \cite{PINH03}.

The direct demonstration of our result can be found in Ref.\ \cite{THIB97}. However, it uses the approach of Refs. \cite{TSAN80,TSAN81} and requires a great deal of preliminary work besides being particularly tedious. The calculation goes in three steps. \emph{Step I:} compute in real space the Green's function $G$ of the electric field in a medium containing a single sphere, for arbitrary positions of the source and observer points, inside or outside the sphere. This Green's function is obtained as a sum of four complementary parts, each one addressing a typical situation for the emission and observation points, which can be independently located inside or outside the scatterer. Each of these parts is expanded on a basis of vector spherical harmonics (VSH) and the elements of this expansion involve multiplicative combinations of spherical Bessel or Hankel functions as is the rule with spherical scatterers. A crucial aspect of this calculation is that it appeals to longitudinal-electric components of the field, as in Ref.\ \cite{TSAN80}, in contrast with older approaches to the problem of wave scattering by spheres \cite{TAI71} where these components are ignored. They have since been recognized as playing an important part in source regions \cite{CHEW90}, being responsible for evanescent modes originating from the scatterers. \emph{Step II} consists in taking bi-variate Fourier transforms of this Green's function with respect to the source and observer positions, before applying manipulations that allow one to extract from it the off-shell T-matrix, see equation (\ref{eq:extract}) where $G_m$ denotes the dipolar Green's function in free space. This step is the most difficult one, since it involves non-trivial definite Fourier integrals on separate ranges, $0<r<a$ and $a<r<\infty$, where $a$ is the sphere radius and $r$ is a radial coordinate.
We could not express these definite integrals in closed form, but instead reduced them to a lengthy sum of explicit terms, added to a residual definite integral of simple form. The latter fortunately cancels out with an identical term arising from the VSH expansion of the Dirac singularity at the origin of $G_m$, to be subtracted from the $G$ in the process of extracting the T-matrix according to equation (\ref{eq:extract}). Hence the result can be expressed in closed form as in Ref.\ \cite{TSAN80}, without any non-evaluated integrals. In \emph{Step III}, some tedious re-organizations of terms are carried out to bring the result into a more usable form that displays all symmetry properties of interest.

The complications makes it problematic to present a concise exposition of this direct approach in the case of dielectric and magnetic contrast, so it will not be pursued here. Instead, we outline hereafter a new and shorter --albeit non-deductive-- proof of the result, which bypasses most of the difficulties of the direct approach. The proof consists in showing that the T-matrix satisfies a defining relation of the T-matrix, namely Eq.\ (\ref{tdef2}). This verification only requires carrying out integrals by a method that can be explained relatively easily. Moreover, these integrals only involve simple poles whose residues can almost be read by inspection. Carrying out such a check remains a cumbersome task, but is a straightforward thing to do with a minimum amount of preliminary technical material.

The paper is organized as follows. Our Fourier transform conventions are explained in Appendix A. After setting up our formalism in Sec.\ \ref{tgfftef}, the coefficients of the Mie series of the off-shell T-matrix are given in Sec.\ \ref{ec}, formulated in such a way that important symmetry and conservation properties \cite{WATE71} (among which the unitarity identity \cite{JOAC75,FITZ95}) are made conspicuous. These properties are discussed in Sections \ref{bp} and \ref{ui}. The principle of a proof of our result is detailed afterwards in Sec.\ \ref{sec:dem}. For purposes of clarity and further physical insight on the structure of the $T$-matrix, we use an intermediate decomposition of the $T$-matrix in intermediate partial ``dielectric'' and ``magnetic'' parts, which originate from similar decompositions of the scattering potential. Before concluding in Sec.\ \ref{c}, some limiting cases of particular interest are examined, and new expressions relevant to applications to spatially-dispersive effective-medium approximations are obtained in Sec.\ \ref{sec:partlim}.

Henceforth, the sign $\times$ stands for the three-dimensional vector product, unless otherwise indicated.

\section{Green's function associated to the electric field and scattering potential operator}
\label{tgfftef}
The $T$-matrix is most easily expressed in the time-harmonic domain and space Fourier representation. The dipolar Green's function ${\sf G}$ associated with the electromagnetic field in an infinite medium of relative isotropic permittivity and permeability $\varepsilon_m$ and $\mu_m$ (the index $m$ referring to the embedding matrix) is the retarded solution of the inhomogeneous wave-propagation equation \cite{CHEW90}
\begin{equation}
\label{defG}
\quad
\left[\mu_m^{-1}{\bm \nabla}\times{\bm \nabla}\times -(\omega /c)^2\varepsilon_m\right] {\sf G}_m(\mathbf{ r}|\mathbf{ r}')={\sf I}\,\delta ^3(\mathbf{ r}-\mathbf{ r}').
\end{equation}
In the Fourier representation, it reads
\begin{equation}
\label{gk}
{\sf G}_m(\mathbf{ k})=\mu_m\left[\frac{{\sf I}-\mathbf{\hat k\hat k}}{k^2-(k_m+i0^+)^2}- \frac{\mathbf{ \hat k\hat k}}{k_m^2}\right],
\end{equation}
where $k_m^2=(\omega/c)^2\varepsilon_m\mu_m$, the angular frequency $\omega$ being considered as a mere parameter hereafter. In Eq.\ (\ref{defG}) the permeability $\mu_m$ is introduced so as to make the source term a pure (permeability-independent) electric current. This set-up allows for a consistent treatment of media with heterogeneous magnetic permeability.

Because of translation invariance, ${\sf G}_m(\mathbf{ r}|\mathbf{ r}')\equiv {\sf G}_m(\mathbf{ r}-\mathbf{ r}')$ so that ${\sf G}_m(\mathbf{ k}|\mathbf{ k}')\equiv {\sf G}_m(\mathbf{ k})\delta(\mathbf{ k}-\mathbf{ k}')$, which defines $\mathsf{G}_m(k)$. In Eq.\ (\ref{gk}), ${\sf I}$ stands for the identity matrix, and $\mathsf{G}_m$ is expressed in terms of the transverse and longitudinal projectors with respect to the direction $\mathbf{\hat k}=\mathbf{ k}/k$ of the Fourier mode $\mathbf{k}$.

Consider now the one-body inhomogeneous problem in presence of a spherical scatterer of radius $a$ centered at the origin. With our above convention for the permeability, the constitutive properties of the medium are specified by
\begin{equation}
\label{constit1}
\varepsilon(\mathbf{ r})=\varepsilon_m+(\varepsilon_s-\varepsilon_m)\phi(\mathbf{ r}),\quad
\frac{1}{\mu(\mathbf{ r})}=\frac{1}{\mu_m}+\left(\frac{1}{\mu_s}-\frac{1}{\mu_m}\right)\phi(\mathbf{ r}),
\end{equation}
where $\varepsilon_s$, $\mu_s$ are the relative permittivity and permeability of the sphere, of characteristic function $\phi(\mathbf{ r})=\theta(a-|r|)$ ($\theta$ denotes the Heaviside step function). The scattering potential operator ${\sf U}$ between points $\mathbf{ r}$ and $\mathbf{ r}^\prime$ is \cite{PELL97b}
\begin{equation}
\label{u1}
{\sf U}(\mathbf{ r}|\mathbf{ r}^\prime)=\delta(\mathbf{ r}-\mathbf{ r}^\prime)\left[{\bm \nabla}'\times\left(\frac{1}{\mu_m}-\frac{1}{\mu_s}\right)\phi(\mathbf{ r}'){\bm \nabla}'\times{}+\left(\frac{\omega}{c}\right)^2\left(\varepsilon_s-\varepsilon_m\right)\,{\sf I}\,
\phi(\mathbf{ r}')\right],
\end{equation}
where the prime denotes a derivative with respect to $\mathbf{r}'$; or in Fourier form \cite{PELL97b}:
\begin{equation}
\label{potfour}
{\sf U}(\mathbf{ k}_1|\mathbf{ k}_2)=\frac{1}{(2\pi)^{3/2}}\Biggl[\left(\frac{1}{\mu_s}-\frac{1}{\mu_m}\right)
\mathbf{ k}_1\times\mathbf{ k}_2\times{}+\left(\frac{\omega}{c}\right)^2 (\varepsilon_{s}-\varepsilon_m)\,{\sf I}\Biggr] \phi(\mathbf{ k}_1-\mathbf{ k}_2).
\end{equation}
The Green's function $\mathsf{G}$ associated to the electric field in the medium now obeys the integro-differential equation
\begin{equation}
\label{su2}
\left[\frac{1}{\mu_m}{\bm \nabla}\times{\bm \nabla}\times
-(\omega /c)^2\varepsilon_m\right] {\sf G}(\mathbf{ r}|\mathbf{ r}')={\sf I}\,\delta ^3(\mathbf{ r}-\mathbf{ r}')+\int\!{\rm d}^3\!r_1\,{\sf U}(\mathbf{ r}|\mathbf{ r}_1){\sf G}(\mathbf{ r}_1|\mathbf{ r}').
\end{equation}
In formal operator notation \cite{TSAN80}, and with the help of ${\sf G}_m$, this equation takes the Lippmann-Schwinger form $G=G_m+G_m U G$.

\section{The T-matrix}
\label{sec:ttm}
\subsection{Definitions and Derivation}
The $T$-matrix operator is introduced so that $G=G_m+G_m T G_m$, hence by definition $T=U(1-G_m U)^{-1}$. This operator inversion is difficult to perform directly. Thus, we computed the $T$-matrix following Tsang and Kong \cite{TSAN80}, \textit{i.e.}\ by first solving the one-body problem in real space for $\mathsf{G}(\mathbf{r}|\mathbf{r}')$ with arbitrary positions of the source $\mathbf{r}$ and of the observation point $\mathbf{r}'$ inside or outside the scatterer; then, by going to Fourier transforms; and eventually by extracting $T$ via the relationship
\begin{equation}
\label{eq:extract}
T=G_m^{-1}(G-G_m)G_m^{-1}.
\end{equation}
The drawbacks of this direct approach have been recalled in the Introduction (see \cite{THIB97} for details). Thus, a much shorter albeit non-deductive proof is provided in Sec.\ \ref{sec:dem}, which consists in proving that the following equation for $T$ holds:
\begin{equation}
\label{tdef2}
T=U+U G_m T.
\end{equation}

The $T$-matrix is obtained as an expansion over vector spherical harmonics defined in Appendix A. We depart from other authors \cite{TSAN80,CHEW90,TAI71} by using the orthonormalized VSH basis $\{\mathbf{N}_{ln},\mathbf{Z}_{ln},\mathbf{X}_{ln}\}$ as found in the book by Cohen-Tannoudji {\em et al.} \cite{COHE87}. The number $l\geq 0$ is the multipole index while $-l\leq n \leq l$ is the angular number. With $\Omega_{\mathbf{ k}}$ as the solid angle in direction $\mathbf{ k}$, $\mathbf{N}_{ln}(\Omega_{\mathbf{ k}})$ has the character of a longitudinal electric component, aligned with $\mathbf{ k}$, whereas $\mathbf{Z}_{ln}$ and $\mathbf{X}_{ln}$ are of transverse electric and magnetic character, respectively, and are orthogonal to $\mathbf{ k}$. Both $\mathbf{Z}_{ln}$ and $\mathbf{X}_{ln}$ are nonzero for $l\geq 1$ only. As recalled in the Introduction, longitudinal terms built on $\mathbf{N}_{ln}$ are necessary in any source region \cite{CHEW90}.

The following variables related to dielectric and magnetic contrast are introduced:
\[
\Delta\varepsilon=(\varepsilon_s/\varepsilon_m)-1,\quad \Delta\mu=(\mu_m/\mu_s)-1,\quad \delta\varepsilon=(\varepsilon_m/\varepsilon_s)\Delta\varepsilon,\quad \delta\mu=(\mu_s/\mu_m)\Delta\mu,
\]
Also, we introduce suitably normalized derivatives of the Ricatti-Bessel and Ricatti-Hankel functions, standard in this context, defined as the product of $x$ by the spherical Bessel of Hankel functions $j_l(x)$ or $h_l^{(1)}(x)$, respectively \cite{JACK75}. They read:
\begin{equation}
\label{eq:qdef}
\varphi_{l,\alpha}\equiv \varphi_l(ak_\alpha)=\frac{[ak_\alpha j_l(ak_\alpha)]'}{j_l(ak_\alpha)},\quad
\varphi^{(1)}_{l,\alpha} \equiv \varphi_l^{(1)}(ak_\alpha)=\frac{[ak_\alpha h^{(1)}_l(ak_\alpha)]'}{h^{(1)}_l(ak_\alpha)}\quad (l\ne 0).
\end{equation}
These $\varphi$ functions simplify the evaluation of our forthcoming results in the static limit where $\omega\to 0$. In the above expressions, index $\alpha$ stands either for $m$, $s$, $k$, $1$ or $2$ depending on the argument $k_{\alpha}$ being $k_m=(\omega/c)(\varepsilon_m\mu_m)^{1/2}$, $k_s=(\omega/c)(\varepsilon_s\mu_s)^{1/2}$, $k$, or the outgoing or incoming momenta $k_1$ or $k_2$ respectively. Finally, let
\begin{equation}
S_{l,\alpha\beta}\equiv\frac{\varphi_{l,\alpha}-\varphi_{l,\beta}}{k_\alpha^2-k_\beta^2},\quad
R_{l,\alpha\beta}\equiv\frac{k_\alpha^2 \varphi_{l,\beta}-k_\beta^2 \varphi_{l,\alpha}}{k_\alpha^2-k_\beta^2},\quad J_{l,12}\equiv\frac{j_l(ak_1)}{ak_1}\frac{j_l(ak_2)}{ak_2}.
\end{equation}
Useful limiting behaviors are $\varphi_l(x)=(l+1)-x^2/(2l+3)+O(x^4)$, so that $S_{l,\alpha\beta}\simeq -a^2/(2l+3)$ and $R_{l,\alpha\beta}\simeq l+1$ when $a\to 0$, and $\varphi^{(1)}_l(x)=-l+x^2/(2l-1)+O(x^4)+O\bigl((ix)^{2l+1}\bigr)$.\footnote{For this reason, slightly different normalizations, in the form of alternative functions $Q_l(x)=\varphi_l(x)/(l+1)$ and $Q_l^{(1)}(x)=-\varphi_l^{(1)}(x)/l$ were used by us in Ref.\ \cite{PELL97b}.}

\subsection{Off-shell components}
\label{ec}
Our VSH expansion of the $T$-matrix, which is the main result of this paper (see Introduction), reads
\begin{equation}
\label{tos}
{\sf T}(\mathbf{ k}_1|\mathbf{ k}_2)=\sum_{\atop{l\ge 0}{\mathbf{ A},\mathbf{ B}=\mathbf{ N},\mathbf{ Z},\mathbf{ X}}}
T_l^{AB}(k_1|k_2)\sum_{n=-l}^l\mathbf{ A}_{ln}(\Omega_{\mathbf{ k}_1})\mathbf{ B}^*_{ln}(\Omega_{\mathbf{ k}_2}),
\end{equation}
where $T_l^{NX}(k_1|k_2)$ $=$ $T_l^{XN}(k_1|k_2)$ $=$ $T_l^{ZX}(k_1|k_2)$ $=$ $T_l^{XZ}(k_1|k_2)$ $=$ $0$ due to spherical symmetry, and
\begin{subequations}
\label{elems}
\begin{eqnarray}
\label{tnn}
&&T^{NN}_l(k_1|k_2)=\frac{2a^3}{\pi}
\frac{k_m^2}{\mu_m}\delta\varepsilon\Biggl[
\frac{l(l+1)\delta\varepsilon}{(\varepsilon_m/\varepsilon_s) \varphi_{l,s}-\varphi^{(1)}_{l,m}}+R_{l,12}-1\Biggr]J_{l,12},\\
\label{tnz}
&&\frac{T^{NZ}_l(k_1|k_2)}{\sqrt{l(l+1)}}=\frac{T^{ZN}_l(k_2|k_1)}{\sqrt{l(l+1)}}=
\frac{2a^3}{\pi}\frac{k_m^2}{\mu_m}
\delta\varepsilon\Biggl[\frac{\delta\varepsilon R_{l,2s}+\delta\mu k_2^2 S_{l,2s}}
{(\varepsilon_m/\varepsilon_s)\varphi_{l,s}-\varphi^{(1)}_{l,m}}+1\Biggr]
J_{l,12},\\
\label{tzz}
&&T^{ZZ}_l(k_1|k_2)=\frac{2a^3}{\pi}\frac{k_m^2}{\mu_m}
\Biggl[\frac{(\delta\varepsilon R_{l,1s}+\delta\mu k_1^2 S_{l,1s})(\delta\varepsilon R_{l,2s}+\delta\mu k_2^2 S_{l,2s})}{(\varepsilon_m/\varepsilon_s)\varphi_{l,s}-\varphi_{l,m}^{(1)}}\nonumber\\
&&\hspace{4cm}{}+\delta\mu\frac{k_1^2 k_2^2}{k_m^2}\frac{(k_1^2-k_m^2)S_{l,1s}-(k_2^2-k_m^2)S_{l,2s}}{k_1^2-k_2^2}\nonumber\\
&&\hspace{4cm}{}+\frac{\delta\varepsilon}{k_m^2}\frac{k_2^2(k_1^2-k_m^2)R_{l,1s}-k_1^2(k_2^2-k_m^2)R_{l,2s}}
{k_1^2-k_2^2}\Biggr]J_{l,12},\\
\label{txx}
&&T^{XX}_l(k_1|k_2)=\frac{2a^3}{\pi}\frac{k_1 k_2}{\mu_m}
\Biggl[\frac{(\Delta\mu R_{l,1s}+k_m^2\Delta\varepsilon S_{l,1s})(\Delta\mu R_{l,2s}+k_m^2\Delta\varepsilon S_{l,2s})}{(\mu_m/\mu_s)\varphi_{l,s}-\varphi^{(1)}_{l,m}}\\
&&\hspace{-0.5cm}{}-\frac{\mu_s}{\mu_m }(\Delta\mu k_1^2-k_m^2\Delta\varepsilon)(\Delta\mu k_2^2-k_m^2\Delta\varepsilon)\frac{S_{l,1s}-S_{l,2s}}{k_1^2-k_2^2}-(\Delta\mu R_{l,12}+k_m^2\Delta\varepsilon S_{l,12})\Biggr]J_{l,12}.\nonumber
\end{eqnarray}
\end{subequations}
Elements $T_l^{NN}$ are defined for $l\geq 0$, whereas the other types are defined for $l\geq 1$. At the price of additional algebraic manipulations, we checked that in absence of magnetic contrast ($\mu_s=\mu_m$), these compact and symmetric expressions are equivalent to those by Tsang and Kong \cite{TSAN80} (who use non-orthonormalized VSH). The equality of $T^{NZ}_l(k_1|k_2)$ and $T^{ZN}_l(k_2|k_1)$ is a consequence of reciprocity \cite{CHEW90}.

\subsection{Basic properties}
\label{bp}
Briefly, the main properties enjoyed by these matrix elements are as follows. First, $T_l^{NN}$, $T_l^{ZZ}$ and $T_l^{XX}$ are symmetric under interchange of $k_1$ and $k_2$, and are non-singular for all finite values of $(k_1,k_2)$ (possibly complex). With respect to this property, note that at a zero of $j_l(ak_\alpha)$, $\alpha=1,2$, the product $\varphi_{l,\alpha} j_l(ak_\alpha)$ is always finite. In addition, the triple limit $k_1\to 0$, $k_2\to 0$, $\omega\to 0$ is uniquely defined, because the limits commute. Also, the limit
\begin{equation}
\lim_{k_1\to k_m}\lim_{k_2\to k_m} T^{AB}_l(k_1|k_2)=\lim_{k\to k_m} T^{AB}_l(k|k),
\end{equation}
uniquely defines the so-called `on-shell' elements (a concept relevant to transverse components only, see below).

The denominators identify the $NN$, $NZ$, $ZN$ and $ZZ$ terms as electric multipole contributions, and the $XX$ terms as magnetic multipoles. The prefactor $k_1k_2$ in $T_l^{XX}(k_1|k_2)$ is results from the operator $\mathbf{ k}_1\times\mathbf{k}_2\times$ in the potential expressed by Eq.\ (\ref{potfour}). As in classical Mie scattering (i.e., for the on-shell T-matrix, see Sec.\ \ref{sec:onshell}), transverse electric and magnetic polariton resonances  \cite{FUCH92,BOHR83,RUPP82} arise at complex frequencies for which denominators vanish, namely when:
\begin{equation}
\left\{\atop{\varepsilon_m}
{\mu_m}\right. \varphi_l(ak_s)-\left\{\atop{\varepsilon_s}{\mu_s}\right. \varphi^{(1)}_l(ak_m)=0\qquad (l\geq 1).
\end{equation}

\subsection{Unitarity identity as a consistency check}
\label{ui}
In the non-dissipative case, all constitutive parameters and  $\mathsf{U}(\mathbf{r}|\mathbf{r}')$ are real. Imaginary parts of the T-matrix elements, rooted in the outgoing-wave prescription $+i0^+$ in Eq.\ (\ref{gk}) that defines $\mathsf{G}_m(\mathbf{k})$, arise solely from the $\varphi_{l,m}^{(1)}$. The fact that these imaginary parts are separable in the momenta $k_1$ and $k_2$, see Eqs.\ (\ref{tnn})--(\ref{txx}), is deeply connected with the well-known \emph{unitarity identity} \cite{WATE71,JOAC75,FITZ95}. This identity is a generalization of a well-known statement of energy conservation in scattering theory, \cite{KONG00}, wherein the scattering $\sigma_s$ cross-section is equal to the extinction cross-section of the scatterer $\sigma_e$ in the absence of absorption. As relationships (\ref{consen}) below show, this identity is deeply connected with the fact that the sum of the second and third terms enclosed in braces in Eq.\ (\ref{tzz}) or (\ref{txx}) reduce to the left or right factor in the numerator of the first term in the same expressions, when $k_1=k_m$ or $k_2=k_m$, respectively. This provides an easy consistency check for expressions in Eqs.\ (\ref{tnn})--(\ref{txx}). For definiteness and further reference, this unitarity identity is derived in Appendix \ref{sec:appunit} in the vector case. Its generic operator form reads
\begin{equation}
\label{im1}{\frac 1{2i}}\bigl(T-T^{\dagger}\bigr)=T^{\dagger}\,
\text{Im}(G_m)\,T=T\,\text{Im}(G_m)\,T^{\dagger }.
\end{equation}
For spherical scatterers, it takes the form (see Appendix \ref{sec:appunit})
\begin{equation}
\label{im4}
\text{Im}{\sf T}(\mathbf{ k}_1|\mathbf{ k}_2)=\frac \pi 2\mu_m k_m
\int\!{\rm d}\Omega _q\,{\sf T}(\mathbf{ k}_1|k_m\mathbf{\hat q})\left(\,{\sf I}-
\mathbf{\hat q}\mathbf{\hat q}\,\right) {\sf T}^{*}(k_m\mathbf{\hat q}|\mathbf{ k}_2).
\end{equation}
Ensuing identities for matrix components are obtained as follows. Since for a sphere $T^{NX}$ $=$ $T^{XN}$ $=$ $T^{NX}$ $=$ $T^{XN}$ $=$ $0$, the sums over $n$ in Eq.\ (\ref{tos}) are purely real (see Appendix \ref{sec:vsh} for their explicit value). Then ${\sf T}^*(\mathbf{k}_1|\mathbf{k}_2)$ is expressed by Eq.\ (\ref{tos}) provided that $T^{AB}_l(k_1|k_2)$ is replaced by $T^{AB*}_l(k_1|k_2)$ in this expression. Then, expanding identity in Eq.\ (\ref{im4}) on the VSH basis and identifying mutually orthogonal components leads to the following relations, to be obeyed for each $l$:
\begin{subequations}
\label{consen}
\begin{eqnarray}
\label{consenx}
\text{Im} T_l^{XX}(k_1|k_2)&=&\frac{\pi}{2} \mu_m k_m T^{XX}_l(k_1|k_m)T^{XX*}_l(k_m|k_2),\\
\label{consenz}
\text{Im} T_l^{ZZ}(k_1|k_2)&=&\frac{\pi}{2} \mu_m k_m T^{ZZ}_l(k_1|k_m)T^{ZZ*}_l(k_m|k_2),\\
\text{Im} T_l^{ZN}(k_1|k_2)&=&\frac{\pi}{2} \mu_m k_m T^{ZZ}_l(k_1|k_m)T^{ZN*}_l(k_m|k_2),\\
\text{Im} T_l^{NZ}(k_1|k_2)&=&\frac{\pi}{2} \mu_m k_m T^{NZ}_l(k_1|k_m)T^{ZZ*}_l(k_m|k_2),\\
\label{consenn}
\text{Im} T_l^{NN}(k_1|k_2)&=&\frac{\pi}{2} \mu_m k_m T^{NZ}_l(k_1|k_m)T^{ZN*}_l(k_m|k_2).
\end{eqnarray}
\end{subequations}
For $l=0$, the last equation must be replaced by $\text{Im} T_0^{NN}(k_1|k_2)=0$. These relations can be explicitly checked on the matrix elements themselves with the help of the formulas (the second equality stems from the Wronskian \cite{JACK75} $W(j_l,h_l^{(1)})=i/x^2$):
\begin{subequations}
\begin{eqnarray}
\label{constraints}
\text{Im} \varphi_l^{(1)}(x)&=&\bigl[x|h^{(1)}_l(x)|^2\bigr]^{-1},\\
\label{eq:wronskian}
\varphi_l(x)-\varphi^{(1)}_l(x)&=&\bigl[i x h_l^{(1)}(x)j_l(x)]^{-1}.
\end{eqnarray}
\end{subequations}

We close this section with the following remark. Combined with the symmetry properties of the matrix elements, Eqs.\ (\ref{consenx}), (\ref{consenz}) and (\ref{consenn}) imply, for real $k$ and non-dissipative media, the positivity of $\text{Im} T_l^{NN}(k|k)$, $\text{Im} T_l^{ZZ}(k|k)$ and $\text{Im} T_l^{XX}(k|k)$. Moreover, setting $\tau_l(k_1|k_2)=(\pi/2)\mu_m k_m T_l(k_1|k_2)$ where $T_l$ stands for either $T_l^{ZZ}$ or $T_l^{XX}$, it is easily seen that identities in Eqs.\ (\ref{consenx}) and (\ref{consenz}) imply the existence of real symmetric  functions $t_l(k_1|k_2)$ such that
\begin{equation}
\label{tau12}
\tau_l(k_1|k_2)=t_l(k_1|k_2)+2i\frac{t_l(k_1|k_m) t_l(k_m|k_2)}{1+\sqrt{1-4t_l^2(k_m|k_m)}},\qquad
|t_l(k_m|k_m)|\le \frac{1}{2},
\end{equation}
a choice of sign in front of the square root having been made. In words, the real part of $T^{XX}$ or $T^{ZZ}$ fully determines the latter quantities as functions of $k_1$ and $k_2$. The $\omega$-dependence of T is closely tied to its $k$-dependence, as shown by the way $k_m$ occurs in the imaginary part of Eq.\ (\ref{tau12}). Finally, the inequality in Eq.\ (\ref{tau12}) allows one to define phase shifts $\delta_l$ (real in absence of dissipation) different for the $ZZ$ and $XX$ components, such that $\tau_l(k_m|k_m)=\sin(\delta_l)\exp(i\delta_l)$ \cite{JOAC75,NEWT66,OHTA80}. Both $t_l$ and $\delta_l$ are analytically continued as functions of $\varepsilon_s$, $\varepsilon_m$, $\mu_s$ and $\mu_m$ in the dissipative case.

\subsection{Sketch of a proof of Eqs.\ (\ref{elems})}
\label{sec:dem}
We prove Eqs.\ (\ref{elems}) by verifying relationship (\ref{tdef2}), taken as a definition of $T$. To proceed, and anticipating further applications to heterogeneous media with spherical inclusions, it is convenient to split up the scattering potential $U$ into its `dielectric' and `magnetic' parts, $U^\varepsilon$ and $U^\mu$, defined from Eq.\ (\ref{potfour}) by alternatively suppressing the magnetic or the dielectric contrast: $U^\varepsilon=U|_{\mu_s=\mu_m}$ and $U^\mu=U|_{\varepsilon_s=\varepsilon_m}$ \cite{PELL97a}. Our aim is to show explicitly that
\begin{equation}
\label{eq:idshow}
T=(U^\varepsilon+U^\varepsilon G_m T)+(U^\mu+U^\mu G_m T),
\end{equation}
which is equivalent to proving Eq.\ (\ref{tdef2}).

From this perspective, VSH representations of $U^\varepsilon$ and $U^\mu$ are obtained from the $T$-matrix components of Eqs.\ (\ref{elems}) by keeping only their lowest-order term in an expansion in powers of the dielectric and magnetic contrasts $\Delta\varepsilon$ and $\Delta\mu$,\footnote{This requires Taylor-expanding $\delta\varepsilon$ and $\delta\mu$ in powers of these quantities, too.} since $U^{\varepsilon,\mu}$ are proportional to these quantities. The following nonzero elements are obtained:
\begin{subequations}
\label{uelmts}
\begin{eqnarray}
\label{uenn}
&&U^{\varepsilon\,NN}_l(k_1|k_2)=\frac{2a^3}{\pi}
\frac{k_m^2}{\mu_m}\Delta\varepsilon(R_{l,12}-1)\,J_{l,12},\\
\label{uenz}
&&U^{\varepsilon\,NZ}_l(k_1|k_2)=U^{\varepsilon\,ZN}_l(k_2|k_1)
=\frac{2a^3}{\pi}\sqrt{l(l+1)}\frac{k_m^2}{\mu_m}\Delta\varepsilon\,J_{l,12},\\
\label{uezz}
&&U^{\varepsilon\,ZZ}_l(k_1|k_2)=\frac{2a^3}{\pi}\frac{k_m^2}{\mu_m}\Delta\varepsilon R_{l,12}\,J_{l,12},\\
\label{uexx}
&&U^{\varepsilon\,XX}_l(k_1|k_2)=-\frac{2a^3}{\pi}\frac{k_1 k_2}{\mu_m} k_m^2\Delta\varepsilon S_{l,12}\,J_{l,12},\\
\label{umzz}
&&U^{\mu\,ZZ}_l(k_1|k_2)=\frac{2a^3}{\pi}\frac{k_1^2 k_2^2}{\mu_m}\Delta\mu S_{l,12}\,J_{l,12},\\
\label{umxx}
&&U^{\mu\,XX}_l(k_1|k_2)=-\frac{2a^3}{\pi}\frac{k_1 k_2}{\mu_m} \Delta\mu R_{l,12}\,J_{l,12}.
\end{eqnarray}
\end{subequations}
Next, one must compute $U^\varepsilon G_mT$ and $U^\mu G_mT$. The calculation goes as follows: first, expand for instance $\int {\sf U}^\varepsilon(\mathbf{ k}_1|\mathbf{ q}_1){\sf G}_m(\mathbf{ q}_1|\mathbf{ q}_2){\sf T}(\mathbf{ q}_2|\mathbf{ k}_2)\, {\rm d}^3\!q_1\,{\rm d}^3\!q_2$ on the VSH basis. Using the orthonormalization properties, we arrive at expressions such as
\begin{eqnarray*}
(U^\varepsilon G_mT)^{ZZ}_l(k_1|k_2)&=&4\pi\int_0^\infty{\rm d}q\,q^2\Bigl[g^T(q) U^{\varepsilon\,ZZ}_l(k_1|q) T^{ZZ}_l(q|k_2)\nonumber\\
&&\hspace{2cm}{}+ g^L(q) U^{\varepsilon\,ZN}_l(k_1|q)T_l^{NZ}(q|k_2) \Bigr],
\end{eqnarray*}
where $g^T$ and $g^L$ are the transverse and longitudinal parts of ${\sf G}_m$ that we have written ${\sf G}_m(\mathbf{ k})\equiv g^T(k)(\,{\sf I}-\mathbf{\hat k}\mathbf{\hat k}\,)+g^L(k)\mathbf{\hat k}\mathbf{\hat k}$. Such expressions only involve products of the form $ g^T(q) U^{\varepsilon\,AZ}_l(k_1|q) T^{ZB}_l(q|k_2)$, or $g^T(q) U^{\varepsilon\,AX}_l(k_1|q) T^{XB}_l(q|k_2)$ or $g^L(q) U^{\varepsilon\,AN}_l(k_1|q) T^{NB}_l(q|k_2)$, where $A,B=N,Z,X$. Upon going back to definition (\ref{eq:qdef}) of the function $\varphi_l(x)$, one observes that these integrals over $q$ all reduce to generic contributions of the type
\begin{equation*}
I=\int_0^\infty \hspace{-1ex}{\rm d}q f(q^2) \Bigl\{ \alpha(q^2)j_l(aq)+\beta(q^2)[\,aq\,j_l(aq)\,]'\Bigr\}\Bigl\{ \gamma(q^2)j_l(aq)+\delta(q^2)[\,aq\,j_l(aq)\,]' \Bigr\},
\end{equation*}
where $f$, $\alpha$, $\beta$, $\gamma$ and $\delta$ are rational functions of $q^2$. These integrals are computed by the following standard method in presence of trigonometric or Bessel functions \cite{WATS80}. Splitting up $I$ into two equal parts by writing $I=I/2+I/2$, then performing alternatively the substitution $j_l(aq)=[h_l^{(1)}(aq)+h_l^{(2)}(aq)]/2$ in the right-hand factor (in the first instance of $I/2$) and in the left-hand one (in the second instance), and appealing next to the change of variable $q\rightarrow -q$, with $h_l^{(2)}(-aq)$ $=$ $(-1)^lh_l^{(1)}(aq)$, yields the equivalent form
\begin{eqnarray}
I&=&\frac{1}{4}\int_{-\infty}^\infty {\rm d}q f(q^2)\Bigl(\Bigl\{ \alpha\,j_l(aq)+\beta\,[\,aq\,j_l(aq)\,]' \Bigr\}\Bigl\{ \gamma\,h_l^{(1)}(aq)+\delta\,[\,aq\,h_l^{(1)}(aq)\,]' \Bigr\}\nonumber\\
\label{eq:integrali}
&&\hspace{1cm}{}+\Bigl\{ \alpha\,h_l^{(1)}(aq)+\beta\,[\,aq\,h_l^{(1)}(aq)\,]' \Bigr\}\Bigl\{ \gamma\,j_l(aq)+\delta\,[\,aq\,j_l(aq)\,]' \Bigr\}\Bigl).
\end{eqnarray}
To be precise, we indicate that this transformation turns products $R_{l,1q}\, j_l(aq)$ and $S_{l,1q}\, j_l(aq)$ in the original integral into the following new quantities:
\vspace{-5mm}
\begin{subequations}
\begin{eqnarray}
\label{eq:ex1}
R_{l,1q}\, j_l(aq)&\to&\frac{q^2\varphi_{l,1}-k_1^2\varphi^{(1)}_{l,q}}{q^2-k_1^2}h^{(1)}_l(aq),\\
\label{eq:ex2}
S_{l,1q} j_l(aq)&\to&\frac{\varphi^{(1)}_{l,q}-\varphi_{l,1}}{q^2-k_1^2}h^{(1)}_l(aq).
\end{eqnarray}
\end{subequations}
Once cast in the form of Eq.\ (\ref{eq:integrali}) the integral can be computed by contour integration, closing the integration path on the real axis using a half-circle of infinite radius in the upper half-plane. Since the transformation generates products of $j_l$ and $h^{(1)}_l$ functions, the contribution of this half-circle vanishes. Besides poles $\pm k_m$ due to the transverse part of the Green's function (if present), the transformation endows the integrand with a single pole at $q=0$ due to products $j_l(aq) h_l^{(1)}(a q)$, and double poles among $\pm k_s$, $\pm k_1$ or $\pm k_2$ because the functions $R$ and $S$ have been modified according to Eqs.\ (\ref{eq:ex1}), (\ref{eq:ex2}). The pole $q=0$ must be handled by a principal value prescription, whereas in the pairs of poles of opposite sign that of minus (resp., plus) sign is shifted in the lower (resp., upper) half-plane by an infinitesimal amount. In this way only poles with plus sign contribute. Most often in these integrals the associated residues can be read by inspection. Extensive use is made of the Wronskian identity (\ref{eq:wronskian}) in subsequent reorganizations to reduce residue contributions coming from terms such as Eqs.\ (\ref{eq:ex1}) or (\ref{eq:ex2}). For instance, the pole $q=k_1$ generated by (\ref{eq:ex2}) gives rise to a residue proportional to $(\varphi_{l,1}- \varphi^{(1)}_{l,1})h^{(1)}_l(ak_1)$, equal to $1/[i a k_1 j_l(ak_1)]$ by virtue of Eq.\ (\ref{eq:wronskian}). In general, the remaining function $j_l(ak_1)$ in this denominator cancels with a similar factor present in the numerator of the multiplying term within braces in the integrand of Eq.\ (\ref{eq:integrali}), in which the $j_l$ have not been transformed. Albeit lengthy, the calculation is thus straightforward.

Adding, respectively, the contributions of potentials $U^\varepsilon$, $U^\mu$ read from Eq.\ (\ref{uelmts}) to the matrix elements of $U^\varepsilon G_mT$ and $U^\mu G_mT$ computed by this procedure yields:
\begin{subequations}
\begin{eqnarray}
\label{uetnn}
&&
\left(U^\varepsilon+U^\varepsilon G_m T\right)^{NN}_l\hspace{-0.4em}(k_1|k_2)=T^{NN}_l\hspace{-0.2em}(k_1|k_2),
\\
\label{uetnz}
&&
\left(U^\varepsilon+U^\varepsilon G_mT\right)^{NZ}_l\hspace{-0.4em}(k_1|k_2)=
\left(U^\varepsilon+U^\varepsilon G_mT\right)^{ZN}_l\hspace{-0.4em}(k_2|k_1)=T^{NZ}_l\hspace{-0.2em}(k_1|k_2),
\\
\label{uetzz}
&&\left(U^\varepsilon+U^\varepsilon G_mT\right)^{ZZ}_l\hspace{-0.4em}(k_1|k_2)=\frac{2a^3}{\pi}
\frac{k_m^2}{\mu_m}\Biggl[\frac{\delta\varepsilon R_{l,1s}(\delta\varepsilon R_{l,2s}+\delta\mu k_2^2 S_{l,2s})}
{(\varepsilon_m/\varepsilon_s)\varphi_{l,s}-\varphi_{l,m}^{(1)}}\nonumber\\
&&\hspace{3.5cm}{}+\frac{\delta\varepsilon}{k_m^2}\frac{k_2^2(k_1^2-k_m^2)R_{l,1s}-k_1^2(k_2^2-k_m^2)R_{l,2s}}
{k_1^2-k_2^2}\Biggr]J_{l,12},
\\
\label{uetxx}
&&\left(U^\varepsilon+U^\varepsilon G_mT\right)^{XX}_l\hspace{-0.4em}(k_1|k_2)
=\frac{2a^3}{\pi}\frac{k_1 k_2}{\mu_m}\Biggl[\frac{k_m^2\Delta\varepsilon S_{l,1s}(\Delta\mu R_{l,2s}+k_m^2\Delta\varepsilon S_{l,2s})}{(\mu_m/\mu_s)\varphi_{l,s}-\varphi^{(1)}_{l,m}}\nonumber\\
&&\hspace{1.7cm}{}+\frac{\mu_s}{\mu_m }k_m^2\Delta\varepsilon(\Delta\mu k_2^2-k_m^2\Delta\varepsilon)\frac{S_{l,1s}-S_{l,2s}}{k_1^2-k_2^2}-k_m^2\Delta\varepsilon S_{l,12}\Biggr]J_{l,12},
\\
\label{umtzz}
&&\left(U^\mu+U^\mu G_mT\right)^{ZZ}_l\hspace{-0.4em}(k_1|k_2)=\frac{2a^3}{\pi}\frac{k_m^2}{\mu_m}
\Biggl[\frac{\delta\mu k_1^2 S_{l,1s}(\delta\varepsilon R_{l,2s}+\delta\mu k_2^2 S_{l,2s})}
{(\varepsilon_m/\varepsilon_s)\varphi_{l,s}-\varphi_{l,m}^{(1)}}\nonumber\\
&&\hspace{3.5cm}{}+\delta\mu\frac{k_1^2 k_2^2}{k_m^2}\frac{(k_1^2-k_m^2)S_{l,1s}-(k_2^2-k_m^2)S_{l,2s}}{k_1^2-k_2^2}\Biggr]J_{l,12},
\end{eqnarray}
\vspace{-1cm}
\begin{eqnarray}
\label{umtxx}
&&\left(U^\mu +U^\mu G_mT\right)^{XX}_l\hspace{-0.4em}(k_1|k_2)=\frac{2a^3}{\pi}\frac{k_1 k_2}{\mu_m}
\Biggl[\frac{\Delta\mu R_{l,1s}(\Delta\mu R_{l,2s}+k_m^2\Delta\varepsilon S_{l,2s})}
{(\mu_m/\mu_s)\varphi_{l,s}-\varphi^{(1)}_{l,m}}\nonumber\\
&&\hspace{2cm}{}-\frac{\mu_s}{\mu_m}\Delta\mu k_1^2(\Delta\mu k_2^2-k_m^2\Delta\varepsilon)\frac{S_{l,1s}-S_{l,2s}}{k_1^2-k_2^2}-\Delta\mu R_{l,12}\Biggr]J_{l,12},
\end{eqnarray}
\end{subequations}
other matrix elements being zero. Identity (\ref{eq:idshow}) can now be checked by mere inspection by comparing these expressions to the matrix elements in Eqs.\ (\ref{elems}).

\section{Limits and values of interest}
\label{sec:partlim}
Some particular limits and values of interest are now examined. The sphere volume is $v=(4\pi/3) a^3$.
\subsection{Point-like limit}
In the mathematical ``point-like" limit where the sphere radius $a$ goes to zero, the matrix elements in (\ref{elems}) reduce to
\begin{subequations}
\begin{eqnarray}
T^{NN\,\rm pt}_l(k_1|k_2)&=&\frac{4\pi v}{(2\pi)^3}\left(\frac{\omega}{c}\right)^2 \varepsilon_m\frac{\varepsilon_s-\varepsilon_m}{\varepsilon_s+2\varepsilon_m}\delta_{l1},\\
T^{XX\,\rm pt}_l(k_1|k_2)&=&\frac{8\pi v}{(2\pi)^3}\frac{k_1 k_2}{\mu_m}\frac{\mu_s-\mu_m}{\mu_s+2\mu_m}\delta_{l1}.
\end{eqnarray}
\end{subequations}
and $T^{NZ\,\rm pt}_l(k_1|k_2)=T^{ZN\,\rm pt}_l(k_1|k_2)=\sqrt{2}T^{NN\,\rm pt}_l(k_1|k_2)$ while
$T^{ZZ\,\rm pt}_l(k_1|k_2)=2 T^{NN\,\rm pt}_l(k_1|k_2)$. To lowest order in the sphere radius, the T-matrix thus reads
\begin{equation}
{\sf T}^{\rm pt}(\mathbf{k}_1|\mathbf{k}_2)=\frac{1}{(2\pi)^{3}}\left[ (\omega/c)^{2}\varepsilon_m\alpha_{\varepsilon}{\sf I} - \frac{\alpha_\mu}{\mu_m}\mathbf{k}_1\times\mathbf{k}_2 \times\ \,\right],
\label{tpoint}
\end{equation}
where $\alpha_\mu$ and $\alpha_\varepsilon$ are the quasi-static electric and magnetic polarizabilities of a sphere \cite{JACK75}:
\begin{equation}
\alpha_\varepsilon=4\pi a^{3}\frac{\varepsilon_s-\varepsilon_m}{\varepsilon_s + 2\varepsilon_m}
 \qquad
\alpha_\mu=4\pi a^{3}\frac{\mu_s-\mu_m}{\mu_s+2\mu_m}
\end{equation}
The quasi-static expression for electric polarizability is too crude to obey the unitarity identity, and a variety of prescriptions have been developed in recent years to include finite frequency corrections to the point-like model that satisfy unitarity \cite{LAGE96}. Corrections to the quasi-static limit that satisfy both unitarity and causality were developed in Ref.\ \cite{PELL97b}.

\subsection{Transverse on-shell elements}
\label{sec:onshell}
For scatterers immersed in a homogenous background media, the calculation of physical quantities proceeds via T-matrices sandwiched between the homogeneous media Green's function $G_m$ (\textit{e.g.}, $G_m T G_m$), the poles of which select the ``on-shell'' T-matrix elements with $k_1=k_2=k_m$. In the scattering and extinction cross-section calculations of Eq.\ (\ref{theopt}), the on-shell T-matrix elements are proportional to the Mie coefficients classically obtained by solving the exterior problem where the source and the observer both lie outside the sphere \cite{BOHR83}. Specialization to this case of expressions (\ref{tzz}), (\ref{txx}) after a few reorganizations that involve the Wronskian identity (\ref{eq:wronskian}), yields the standard values of these coefficients, which in our notations reads (see also \cite{PELL97b})
\begin{subequations}
\label{tonsh}
\begin{align}
\label{tzzmm}
\hspace{-1em}i\frac{\pi}{2}\mu_m k_m T_{l}^{ZZ}(k_m|k_m) &
=\frac{\,j_{l}(ak_m)}{h_{l}^{(1)}(ak_m)}\frac{\varepsilon_{s}\varphi_{l,m}-\varepsilon_m \varphi_{l,s}}{\varepsilon_m \varphi_{l,s}-\varepsilon_{s}\varphi_{l,m}^{(1)}},\\
\label{txxmm}
\hspace{-1em}i\frac{\pi}{2}\mu_m k_m T_{l}^{XX}(k_m|k_m)  &
=\frac{\,j_{l}(ak_{m})}{h_{l}^{(1)}(ak_m)}\frac{\mu_{s}\varphi_{l,m}-\mu_m
\varphi_{l,s}}{\mu_m \varphi_{l,s}-\mu_{s}\varphi_{l,m}^{(1)}}.
\end{align}
\end{subequations}
The right hand sides of these equations are the dimensionless T-matrix elements typically manipulated in on-shell theories. The factor $\pi\mu_m k_m/2$ arises from slightly different conventions and normalizations that are generally practiced between off and on-shell theories.

\subsection{Equal momenta}
\label{sec:eqmom}
The case of forward scattering $\mathbf{k}_1=\mathbf{k}_2=\mathbf{k}$ is particularly important for applications to random media, since it is the one relevant to the computation of the first correction in the volume density of scatterers, to the non-local effective permittivity and permeability of the medium \cite{THIB97,BARR07}. In this case, the Mie series that defines $T$ can be partially re-summed. Only the result is presented here, the calculation is carried out in Appendix \ref{somaparti}.

Symmetry considerations allow us to decompose $\mathsf{T}(\mathbf{k}|\mathbf{k})$ into longitudinal and transverse parts as
\begin{equation}
{\sf T}(\mathbf{ k}|\mathbf{ k})=T^L(k)\, \mathbf{\hat k}\mathbf{\hat k}+T^T(k)\,(\,\mathsf{I}-\mathbf{\hat k}\mathbf{\hat k}\,),
\end{equation}
where $T^T(k)=T^Z(k)+T^X(k)$, and where $T^L$, $T^Z$ and $T^X$ are defined by:
\begin{subequations}
\begin{eqnarray}
&&\sum_{l\geq 0}
T_l^{NN}(k_1|k_2)\sum_{n}\mathbf{ N}_{l n}(\Omega_{\mathbf{ k}_1})\mathbf{ N}_{l n}^*(\Omega_{\mathbf{ k}_2})\equiv T^L(k)\,\mathbf{\hat k}\mathbf{\hat k},\\
&&\sum_{l\geq 1}T_l^{ZZ}(k|k)\sum_{n}\mathbf{ Z}_{l n}(\Omega_{\mathbf{ k}})\mathbf{ Z}_{l n}^*(\Omega_{\mathbf{ k}})\equiv T^Z(k)(\,{\sf I}-\mathbf{\hat k}\mathbf{\hat k}\,),\\
&&\sum_{l\geq 1}
T_l^{XX}(k|k)\sum_{n}\mathbf{ X}_{l n}(\Omega_{\mathbf{ k}})\mathbf{ X}_{l n}^*(\Omega_{\mathbf{ k}})\equiv T^X(k)(\,{\sf I}-\mathbf{\hat k}\mathbf{\hat k}\,).
\end{eqnarray}
\end{subequations}
The sums over $n$ are given by formulas (\ref{eq:sumkk}), from which we deduce that:
\begin{subequations}
\label{tp}
\begin{eqnarray}
\label{tparkperxzk}
T^L(k)=\sum_{l\geq 0}\frac{(2l+1)}{4\pi} T^{NN}_l(k|k),\quad
T^{\left\{\atop{Z}{X}\right.}(k)=\sum_{l\geq 1}\frac{(2l+1)}{8\pi}T^{\left\{\atop{ZZ}
{XX}\right.}_l(k|k)
\end{eqnarray}
\end{subequations}
where the matrix elements, read in Eqs.\ (\ref{elems}) at unequal momenta, can by evaluated in the limit $k_1, k_2\to k$ by means of Eqs.\ (\ref{limites}).
Introduce now the function
\begin{equation}
\label{firstsum}
{\cal S}(x)\equiv\frac{3}{2}\sum_{l\geq 1}(2l+1)\varphi_l(x)\left[\frac{j_l(x)}{x}\right]^2=3\frac{1-j_0(2x)}{2 x^2},
\end{equation}
which is such that ${\cal S}(x)=1-\frac{1}{5}x^2+O(x^4)$. Appendix B shows how part of the sums over $l$ that result from the above limiting process can be expressed using $\mathcal{S}$. One ends up with:
\begin{subequations}
\label{keg}
\begin{eqnarray}
\label{kegpar}
&&\frac{(2\pi)^3}{v}T^L(k)=\frac{k_m^2}{\mu_m} \delta\varepsilon\Biggl\{\sum_{l\geq 1}\frac{3l(l+1)(2l+1)\Delta\varepsilon}{\varphi_{l,s}-(\varepsilon_s/\varepsilon_m)  \varphi^{(1)}_{l,m}}\left[\frac{j_l(a k)}{a k}\right]^2+1\Biggr\},\\
&&\frac{(2\pi)^3}{v}T^{Z}(k)=\frac{k_m^2}{\mu_m}\left[\delta\varepsilon
+\frac{\varepsilon_s}{\varepsilon_m}\frac{(\delta\varepsilon-\delta\mu)^2 k^4}
{(k^2-k_s^2)^2}\right]{\cal S}(ak)\nonumber\\
&&\hspace{3cm}{}+
\frac{1}{2\mu_m}(\delta\mu k^2-\delta\varepsilon k_s^2)\left(\frac{k^2-k_m^2}
{k^2-k_s^2}\right)[{\cal S}(ak)-1]\\
&&{}+\frac{3}{2}\frac{k_m^2}{\mu_m}\frac{\varepsilon_s}{\varepsilon_m}\sum_{l\geq 1}(2l+1)
\Biggl[\frac{(\delta\varepsilon R_{l,ks}+\delta\mu k^2 S_{l,ks})^2}
{\varphi_{l,s}-(\varepsilon_s/\varepsilon_m)\varphi_{l,m}^{(1)}}-\frac{(\delta\varepsilon-\delta\mu)^2 k^4}
{(k^2-k_s)^2}\varphi_{l,s}\Biggr]\left[\frac{j_l(ak)}{ak}\right]^2,\nonumber\\
&&\frac{(2\pi)^3}{v}T^{X}(k)=\frac{k^2}{\mu_m}\left[-\Delta\mu+ \frac{\mu_s}{\mu_m} \frac{(\Delta\mu k^2-\Delta\varepsilon k_m^2)^2}{(k^2-k_s^2)^2}\right]{\cal S}(ak)\nonumber\\
&&\hspace{3cm}{}+\frac{1}{2\mu_m}(\delta\mu k^2-\delta\varepsilon k_s^2)\left(\frac{k^2-k_m^2}
{k^2-k_s^2}\right)[{\cal S}(ak)-1]\\
&&\hspace{3cm}{}+\frac{3}{2}\frac{k^2}{\mu_m}\frac{\mu_s}{\mu_m }\sum_{l\geq 1}(2l+1)\Biggl[\frac{(\Delta\mu R_{l,ks}+k_m^2\Delta\varepsilon S_{l,ks})^2}{\varphi_{l,s}-(\mu_s/\mu_m)\varphi^{(1)}_{l,m}}\nonumber\\
&&\hspace{7cm}{}-\frac{(\Delta\mu k^2-k_m^2\Delta\varepsilon)^2}
{(k^2-k_s^2)^2}\varphi_{l,s}\Biggr]\left[\frac{j_l(ak)}{ak}\right]^2.\nonumber
\end{eqnarray}
\end{subequations}
Expression (\ref{kegpar}) is the frequency-dependent counterpart of the static momentum-dependent expression obtained by Diener and K\"aseberg \cite{DIEN76}, to which it reduces when $\omega\to 0$ (see also Eq.\ (48) of Ref.\ \cite{BARR07}). Apart from the occurrence of different magnetic permeabilities in $k_m$ and $k_s$ that enter the definitions of $\smash{\varphi^{(1)}_{l,m}}$ and $\varphi_{l,s}$, this longitudinal term has the same form as in the case with no magnetic contrast. In Ref.\ \cite{BARR07}, the expression of $T^T(k)$ provided in the case $\mu_m=\mu_s$ involves integrals that are left unevaluated. Instead, the present result is fully explicit: the transverse part at $\mu_s=\mu_m$ follows from using this equality and setting $\delta\mu=\Delta\mu=0$ in the above expressions.

Though this is not obvious from the above, expressions of $T^{Z}(k)$ and $T^{X}(k)$ are regular in the limit $k\to k_s$. This can be shown by using Taylor expansions, more particularly expansion (\ref{devlop}). In this case, it is actually easier to check regularity term-by-term in each individual term of the non-resummed Mie series, see Eqs.\ (\ref{eq:kzegzz}) and (\ref{eq:kzegxx}), to which one can always go back in case of problems in numerical evaluations near this limit.

It should finally be noted that the right-hand side of Eq.\ (\ref{kegpar}) goes to infinity in the limit where $\varepsilon_s\to 0$ (which is almost the case at the plasma frequency in the high-frequency limit of dielectric response \cite{JACK75}), unless $k=0$. Then indeed
\begin{eqnarray}
&&T^L(k)\simeq \frac{v}{(2\pi)^3}\frac{k_m^2}
{\mu_m}\frac{\varepsilon_m}{\varepsilon_s}\left\{3\sum_{l\geq 1} l(2l+1)\left[\frac{j_l(a k)}
{ak}\right]^2-1\right\}\nonumber\\
\label{singes0}
&=&\frac{v}{(2\pi)^3}\frac{k_m^2}{\mu_m}\frac{\varepsilon_m}{\varepsilon_s}\left\{\frac{3}
{4(ak)^2}\left[2ak\mathop{\text{Si}}(2ak)+j_0(2ak)+\cos(2ak)-2\right]-1\right\},
\end{eqnarray}
and the function within braces, which arises from formulas taken from \cite{DIEN76} and where $\mathop{\text{Si}}(x)$ is the sine-integral function, has no other real zero than $k=0$ near which it behaves as $-(ak)^2/15$.

\section{Concluding remarks}
\label{c}
We derived the off-shell T-matrix of a dielectric and magnetic sphere, provided relatively simple means to check this result, and some particular limits of physical importance were examined. Leaving applications to further work, we close with the following remarks.

First, the introduction of the intermediate functions $\varphi_l$, $R_l$ and $S_l$ was found to be a quite useful device in trying to put some order in the structure of our results, and in helping displaying physical symmetries of interest.

Second, it is observed that only the magnetic extension allows one to recover the limiting case of perfectly conducting inclusions:  as is explained in Ref.\ \cite{KONG00} (p.\ 790), this ideal case corresponds to formally taking the {\em joint} limit $\varepsilon_s\to\infty$ and $\mu_s\to 0$ in the scatterer. Such limiting values allow one to retrieve from Eqs.\ (\ref{sex}), and (\ref{tzzmm}), (\ref{txxmm}) the well-known Mie-Debye low-frequency scattering cross-section of a perfectly conducting sphere obtained from Leontovich's boundary condition with surface impedance $Z=0$ (\textit{e.g.}, Ref.\ \cite{JACK75}, formula 16.159). This cross-section is larger by a factor $1.25$ than that found for $\varepsilon_s=\infty$ but $\mu_s=\mu_m$. Similar limits can easily be taken in the off-shell expressions.

It should be remarked that even though the most useful physical quantities are obtained from either on-shell matrix elements, $k_1=k_2=k_m$ (\textit{e.g.}, cross sections) or forward scattering, $\mathbf{ k}_1=\mathbf{ k}_2=\mathbf{ k}$, for effective-medium approaches in random media, it was only by computing first the T-matrix at unequal momenta that we can currently reach in explicit form these quantities of interest. This should be clear from the definition $T=U+U G_m T$, that involves an integration over arbitrary momenta. In the purely dielectric case, an alternative method has been recently proposed \cite{BARR07} to  directly obtain the relevant elements at equal momenta, but the outcome involves integrals to be done numerically, and the method has not yet been extended to magnetic contrast. In this respect, an appealing perspective might consist in comparing our results in absence of magnetic contrast to that of Ref.\ \cite{BARR07} to the purpose of deriving identities for these integrals. This might ultimately lead to a shorter path to obtaining T-matrices at equal momenta in other cases of interest beyond dia- or paramagnetism.

Finally, the behavior in the limit $\varepsilon_s\to 0$ emphasized at the end of the previous section indicates that in this case for $k\not=0$, the perturbative approach that consists in computing the effective longitudinal dispersion relation of a composite medium to one-body order \cite{BARR07} would fail, since the longitudinal part of the T-matrix goes to infinity. Singularities also arise at polariton resonances. In situations of the sort, it has sometimes been found that in effective constitutive parameters, the first correction to the homogeneous matrix changes its usual proportionality to $f$, the volume fraction of inclusions, into a proportionality to some lesser power of $f$ (e.g., \cite{WILL08}). Such cases therefore deserve special attention when considering applications of the present results to effective-medium theories.

\appendices
\section{Fourier transform conventions}
\label{sec:fourconv}
Our Fourier transform conventions are as follows. This work makes use of generic operators, say $A(\mathbf{r}|\mathbf{r}')$, which may contain derivatives, with ``input'' point $\mathbf{r}'$ and output point $\mathbf{r}$. By convention, their space Fourier transform is taken up by multiplying on the right by a factor $e^{-i\mathbf{k}\cdot\mathbf{r}}/(2\pi)^{3/2}$, and on the left by $e^{+i\mathbf{k}'\cdot\mathbf{r}'}/(2\pi)^{3/2}$, and by carrying out the integrals over $\mathbf{r}$ and $\mathbf{r}'$ in the infinite volume to obtain the transform $A(\mathbf{k}|\mathbf{k}')$. This use of a normalized plane-wave basis is standard when dealing with operators.

However, whenever $A(\mathbf{r}|\mathbf{r}')\equiv A(\mathbf{r}-\mathbf{r}')$ is translation invariant (we use the same $A$ by abuse of notation), we write $A(\mathbf{k}|\mathbf{k}')=\delta(\mathbf{k}-\mathbf{k}')A(\mathbf{k})$, which follows from computing $A(\mathbf{k})$ as the transform of the one-entry function $A(\mathbf{r})$ by multiplying the latter by  $e^{-i\mathbf{k}\cdot\mathbf{r}}$ and by integrating over $\mathbf{r}$. This is the standard practice of solid-state physics.

In the present context, this use of two conventions is necessary to spare us from dragging factors $(2\pi)^{3/2}$ in translation-invariant expressions of interest expressed as Fourier transforms. No confusion will result since the use of the operator convention is indicated by the vertical bar between two variables.

\section{Vector spherical harmonics}
\label{sec:vsh}
The Vector Spherical Harmonics used in this work are defined for $l\geq 0$ and $-l\leq n\leq l$ as \cite{COHE87}
\begin{subequations}
\begin{eqnarray}
\mathbf{ N}_{ln}(\Omega_\mathbf{ k})&=&\mathbf{\hat k}\, Y_{ln}(\Omega_\mathbf{ k}),\\
\mathbf{ Z}_{ln}(\Omega_\mathbf{ k})&=&\frac{1}{\sqrt{l(l+1)}}\mathbf{\nabla}_{\Omega_\mathbf{ k}}
Y_{ln}(\Omega_\mathbf{ k}),\\
\mathbf{ X}_{ln}(\Omega_\mathbf{ k})&=&\frac{1}{\sqrt{l(l+1)}}\mathbf{\hat k}\times\mathbf{\nabla}_{\Omega_\mathbf{ k}} Y_{ln}(\Omega_\mathbf{ k}).
\end{eqnarray}
\end{subequations}
where the $Y_{ln}(\Omega)$ are the usual scalar spherical harmonics \cite{JACK75}, and where $\mathbf{\nabla}_{\Omega_{\mathbf{ k}}}$ is the angular part of the differential operator
$\mathbf{\nabla}=\mathbf{\hat k}\,\partial/\partial k+(1/k)\mathbf{\nabla}_{\Omega_{\mathbf{ k}}}$ in spherical coordinates.
Another standard notation for the VSHs is $\mathbf{Y}_{ln}^{(0),(e),(m)}$ (see \textit{e.g.}, \cite{NEWT66}). However, the present notation, already employed by us in Ref.\ \cite{PELL97b}, alleviates the need for superscripts.

The VSH are such that
$\mathbf{ X}_{l n}(\Omega_\mathbf{ k})=\mathbf{\hat k}\times
\mathbf{ Z}_{l n}(\Omega_\mathbf{ k})$ and $\mathbf{ Z}_{l n}(\Omega_\mathbf{ k})=-\mathbf{\hat
k}\times\mathbf{ X}_{l n}(\Omega_\mathbf{ k})$. Observe that $\mathbf{ X}_{0 0}$ and $\mathbf{ Z}_{0 0}$ are identically zero. Under parity, $\mathbf{ X}_{l n}(-\mathbf{\hat k})=(-1)^l \mathbf{ X}_{l n}(\mathbf{\hat k})$, $\mathbf{ N}_{l n}(-\mathbf{\hat k})=(-1)^{l-1} \mathbf{ N}_{l n}(\mathbf{\hat k})$ and $\mathbf{ Z}_{l n}(-\mathbf{\hat k})=(-1)^{l-1} \mathbf{ Z}_{l n}(\mathbf{\hat k})$. These VHS are orthonormalized:
\begin{equation}
\label{ortho1}\int\!{\rm d}\Omega_\mathbf{ k} \mathbf{ A}_{l n}^*(\Omega_\mathbf{ k}). \mathbf{ B}_{l^{\prime}n^{\prime}}(\Omega_\mathbf{ k})=\delta_{A,B} \delta_{l,l^{\prime}} \delta_{n,n^{\prime}},
\end{equation}
where $\mathbf{ A}$, $\mathbf{ B}$ stand indifferently for $\mathbf{ N}$, $\mathbf{ X}$ or $\mathbf{ Z}$. The closure relationship reads:
\begin{equation}
\label{ferm1}
\sum_{l n} \bigl\{\mathbf{ N}_{ln}(\Omega_1)\mathbf{ N}^*_{ln}(\Omega_2)+\mathbf{ Z}_{ln}(\Omega_1)\mathbf{ Z}^*_{ln}(\Omega_2)+\mathbf{ X}_{ln}(\Omega_1)\mathbf{ X}^*_{ln}(\Omega_2)\bigr\}=\mathsf{I}\,\delta(\Omega_1-\Omega_2).
\end{equation}
Introducing $u=\mathbf{\hat k}_1\cdot\mathbf{\hat k}_2$ and the Legendre polynomial $P_l(x)$ defined by the generating function $(1-2tx+t^2)^{-1/2}=\sum_{l\geq 0}P_l(x) t^l$,
the following sums are obtained (\textit{e.g.}, \cite{PELL97b}):
\begin{subequations}
\begin{eqnarray}
\label{n1n2}
\sum_n \mathbf{N}_{ln}(\Omega_{\mathbf{k}_1})\mathbf{N}_{ln}^*(\Omega_{\mathbf{k}_2})&=&\frac{2l+1}{4\pi}P_l(u)
\mathbf{\hat k}_1\mathbf{\hat k}_2,\\
\label{n1z2}
\sum_n \mathbf{N}_{ln}(\Omega_{\mathbf{k}_1})\mathbf{Z}_{ln}^*(\Omega_{\mathbf{k}_2})&=&
\frac{2l+1}{4\pi\sqrt{l(l+1)}}P'_l(u)
\mathbf{\hat k}_1(\mathbf{\hat k}_1-u\mathbf{\hat k}_2),\\
\label{z1n2}
\sum_n \mathbf{Z}_{ln}(\Omega_{\mathbf{k}_1})\mathbf{N}_{ln}^*(\Omega_{\mathbf{k}_2})&=&
\frac{2l+1}{4\pi\sqrt{l(l+1)}}P'_l(u)
(\mathbf{\hat k}_2-u\mathbf{\hat k}_1)\mathbf{\hat k}_2,\\
\label{z1z2}
\sum_n \mathbf{Z}_{ln}(\Omega_{\mathbf{k}_1})\mathbf{Z}_{ln}^*(\Omega_{\mathbf{k}_2})&=&
\frac{2l+1}{4\pi l(l+1)}P''_l(u)
(\mathbf{\hat k}_2-u\mathbf{\hat k}_1)(\mathbf{\hat k}_1-u\mathbf{\hat k}_2)\nonumber\\
&+&
\frac{2l+1}{4\pi l(l+1)}P'_l(u)
(\mathsf{I}-\mathbf{\hat k}_1\mathbf{\hat k}_1-\mathbf{\hat k}_2\mathbf{\hat k}_2+u\mathbf{\hat k}_1\mathbf{\hat k}_2),\\
\label{x1x2}
\sum_n \mathbf{X}_{ln}(\Omega_{\mathbf{k}_1})\mathbf{X}_{ln}^*(\Omega_{\mathbf{k}_2})&=&
\frac{2l+1}{4\pi l(l+1)}P''_l(u)
(\mathbf{\hat k}_1\times\mathbf{\hat k}_2)(\mathbf{\hat k}_2\times\mathbf{\hat k}_1)\nonumber\\
&+&\frac{2l+1}{4\pi l(l+1)}P'_l(u)
(u\mathsf{I}-\mathbf{\hat k}_2\mathbf{\hat k}_1).
\end{eqnarray}
\end{subequations}
Since $P_l(1)=1$ and $P'_l(1)=l(l+1)/2$,  the only non-zero sums at equal angles $\Omega_{\mathbf{k}_1}=\Omega_{\mathbf{k}_2}$ are:
\begin{subequations}
\label{eq:sumkk}
\begin{eqnarray}
\sum_n \mathbf{N}_{ln}\mathbf{N}_{ln}^*&=&\frac{2l+1}{4\pi}
\mathbf{\hat k}\mathbf{\hat k},\\
\sum_n \mathbf{Z}_{ln}\mathbf{Z}_{ln}^*&=&
\sum_n \mathbf{X}_{ln}\mathbf{X}_{ln}^*=\frac{2l+1}{8\pi}
(\mathsf{I}-\mathbf{\hat k}\mathbf{\hat k}).
\end{eqnarray}
\end{subequations}

\section{Unitarity identity}
\label{sec:appunit}
Let $A^{^*}$ the complex conjugate of operator $A$; $A^{^T}$ its transpose; and $A^{\dagger }$ its Hermitian conjugate, in the direct or Fourier representations: Since
$A^{^T}_{ij}(\mathbf{ r}_1|\mathbf{ r}_2)=A_{ji}(\mathbf{ r}_1|\mathbf{ r}_2)$,
$A^{^T}_{ij}(\mathbf{ k}_1|\mathbf{ k}_2)=A_{ji}(\mathbf{ k}_1|\mathbf{ k}_2)$,
$A^{\dagger}_{ij}(\mathbf{ r}_1|\mathbf{ r}_2)=A^{^*}_{ji}(\mathbf{ r}_2|\mathbf{ r}_1)$,
and $A^{\dagger}_{ij}(\mathbf{ k}_1|\mathbf{ k}_2)=A^*_{ji}(\mathbf{ k}_2|\mathbf{ k}_1)$,
operators ${}^T$ and ${}^{\dagger }$ commute with Fourier transforms. Our first step is to express by a condition on $U$ the reality of the constitutive parameters. Potential $U$, as a generalized response function, is subject to Onsager's symmetry principle for kinetic coefficients that translates here into the principle of inverse propagation of light (or reciprocity). Assuming the absence of a constant external magnetic field, this reads: ${\sf U}(\mathbf{ r}_1|\mathbf{ r}_2)={\sf U}^{^T}(\mathbf{ r}_2|\mathbf{ r}_1)$, or ${\sf U}(\mathbf{ k}_1|\mathbf{ k}_2)={\sf U}^{^T}(-\mathbf{ k}_2|-\mathbf{ k}_1)$. Meanwhile,
absence of dissipation translates as ${\sf U}(\mathbf{ r}_1|\mathbf{ r}_2)={\sf U}^{^*}(\mathbf{ r}_1|\mathbf{ r}_2)$, or
${\sf U}(\mathbf{ k}_1|\mathbf{ k}_2)={\sf U}^{^*}(-\mathbf{ k}_1|-\mathbf{ k}_2)$. Combining both sets of equalities implies that $U=U^\dagger$. Therefore, $G$ obeys $G_m^{-1}-G^{-1}=U_{\mathbf{ y}}=U_{\mathbf{ y}}^{\dagger }=G_m^{\dagger -1}-G^{\dagger -1}$, so that using $G=G_m+G_m T G_m$ provides:
\begin{equation}
G_m(T+T G_m^{\dagger}T^{\dagger})G_m^{\dagger}=
G_m(T^{\dagger}+T G_m T^{\dagger})G_m^{\dagger}.
\end{equation}
The desired unitarity identity on the $T$-matrix follows \cite{JOAC75,FITZ95}:
$
T-T^{\dagger }=T^{\dagger }(G_m-G_m^{\dagger})T=T(G_m-G_m^{\dagger })T^{\dagger}
$.
Noticing that $G_m^{\dagger}=G_m^{*}$, one ends up with equation (\ref{im1}). With (\ref{gk}), we have for real $k_m$:
\begin{subequations}
\begin{eqnarray}
\text{Im} {\sf G}_m(\mathbf{ k})&=&\frac{\pi}{2}\frac{\mu_m}{k_m}
(\,{\sf I}-\mathbf{\hat k}\mathbf{\hat k}\,)\,\delta (k-k_m),\nonumber\\
\text{Im} {\sf G}_m(\mathbf{ r})&=&\frac{k_m\mu_m}{16 \pi^2}\int\!{\rm d}\Omega_\mathbf{ k}
(\,{\sf I}-\mathbf{\hat k}\mathbf{\hat k}\,) e^{i k_m \mathbf{\hat k}.\mathbf{ r}}.
\end{eqnarray}
\end{subequations}
Since $T_\mathbf{ y}$ also obeys Onsager's principle, relations (\ref{im1}) can be rewritten as
\begin{subequations}
\label{im3}
\begin{eqnarray}
&&\text{Im} {\sf T}(\mathbf{ r}_1|\mathbf{ r}_2)=\int\!{\rm d}^3\!x_1\,d^3\!x_2\,
{\sf T}(\mathbf{ r}_1|\mathbf{ x}_1) \text{Im}\left({\sf G}_m(\mathbf{ x}_1-\mathbf{ x}_2)\right){\sf T}^*(\mathbf{ x}_2|\mathbf{ r}_2),\\
&&\frac{1}{2 i} \left[{\sf T}(\mathbf{ k}_1|\mathbf{ k}_2)
-{\sf T}^*(-\mathbf{ k}_1|-\mathbf{ k}_2)\right]\\
&&\quad =
\frac{\pi}{2}\mu_m k_m \int\!{\rm d}\Omega_q\,{\sf T}(\mathbf{ k}_1|k_m \mathbf{\hat q})(\,{\sf I}-\mathbf{\hat q}\mathbf{\hat q}\,)
{\sf T}^*(-k_m \mathbf{\hat q}|-\mathbf{ k}_2).\nonumber
\end{eqnarray}
\end{subequations}
Scatterers for which the origin of coordinates is a symmetry center obey the property, inherited from $U$, that ${\sf T}(-\mathbf{ k}_1|-\mathbf{ k}_2)={\sf T}(\mathbf{ k}_1|\mathbf{ k}_2)$. Equation (\ref{im3}) then entails Eq.\ (\ref{im4}) in the main text.

To retrieve the unitarity relations in their usual form, let the incident field be of the form $\mathbf{ E}_{\rm i}(\mathbf{ r})=e^{i(\mathbf{ k}_{\rm i}.\mathbf{ r}-\omega t)} \overline{\mathbf{ E}}_{\rm i}$,
with $|\mathbf{ k}_{\rm i}|=k_m$. The scattered field at large distances from the scatterer, $\mathbf{ E}_{\rm s}$, is such that:
\begin{equation}
\frac{\mathbf{ E}_{\rm s}(\mathbf{ r})}{(2\pi)^3\mu_m }=
\int\!\frac{{\rm d}^3\!x\,d^3\!y}{(2\pi)^3\mu_m }\,{\sf G}_m(\mathbf{ r}-\mathbf{ x})
{\sf T}(\mathbf{ x}|\mathbf{ y})\mathbf{ E}_{\rm i}(\mathbf{ y})\simeq
\frac{e^{i(k_m r-\omega t)}}{4\pi r}(\,{\sf I}-\mathbf{\hat k}_f\mathbf{\hat k}_f\,)
{\sf T}(\mathbf{ k}_f|\mathbf{ k}_{\rm i}) \overline{\mathbf{ E}}_{\rm i}
\end{equation}
where $\mathbf{ k}_f=k_m \mathbf{\hat r}$. The total field reads
$
\mathbf{ E}_{\rm tot}(\mathbf{ r})=\mathbf{ E}_{\rm i}(\mathbf{ r})+\mathbf{ E}_{\rm s}(\mathbf{r}).
$
Accordingly, the total complex Poynting vector is the sum of its incident, scattering, and extinction parts:
$
\mathbf{ S}_{\rm tot}(\mathbf{ r})=\mathbf{ E}_{\rm tot}(\mathbf{ r})\times\mathbf{ H}_{\rm tot}^{*}(\mathbf{ r})
=\mathbf{ S}_{\rm i}(\mathbf{ r})+\mathbf{ S}_{\rm s}(\mathbf{ r})+\mathbf{ S}_{e}(\mathbf{ r})
$,
with
$\mathbf{ S}_{\rm i}(\mathbf{ r})=\mathbf{ E}_{\rm i}(\mathbf{ r})\times\mathbf{ H}_{\rm i}^{*}(\mathbf{r})$,
$\mathbf{ S}_{\rm s}(\mathbf{ r})=\mathbf{ E}_{\rm s}(\mathbf{ r})\times\mathbf{ H}_{\rm s}^{*}(\mathbf{r})$, and
$\mathbf{ S}_{\rm e}(\mathbf{ r})=\mathbf{ E}_{\rm s}(\mathbf{ r})\times\mathbf{ H}_{\rm i}^{*}(\mathbf{r})
+\mathbf{ E}_{\rm i}(\mathbf{ r})\times\mathbf{ H}_{{\rm s}}^{*}(\mathbf{r})$.
Denoting the time-average of the real incident Poynting vector by
$
\langle\mathbf{ S}_{\rm i}\rangle(\mathbf{ r})={\frac{1}{2}}\text{Re} \mathbf{ S}_{\rm i}(\mathbf{ r})$,
the scattering and extinction cross-sections are respectively
\begin{equation}
\sigma_{\rm s}=\frac{1}{|\!|\langle\mathbf{ S}_{\rm i}\rangle(\mathbf{ r})|\!|}
\int\limits_{S_\infty}\!{\rm d}S\,\frac{1}{2}\text{Re} \mathbf{ S}_{\rm s}(\mathbf{ r}),\qquad
\sigma_{\rm e}=-\frac{1}{|\!|\langle\mathbf{ S}_{\rm i}\rangle(\mathbf{ r})|\!|}
\int\limits_{S_\infty}\!{\rm d}S\,\frac{1}{2}\text{Re} \mathbf{ S}_{\rm e}(\mathbf{ r}),
\end{equation}
where surface integrals are performed on a sphere whose radius goes to infinity, centered on the scatterer \cite{JACK75}. The outer medium being lossless, we find after some algebra that
\begin{subequations}
\label{theopt}
\begin{eqnarray}
\qquad
\label{ssca}
\sigma_{\rm s}&=& \frac{4\pi^4\mu_m^2}{\overline{E}_{\rm i}^{2}}
\overline{\mathbf{ E}}_{\rm i}^{*}\cdot\int\!{\rm d}\Omega_f\,{\sf T}^{\dagger}(\mathbf{k}_i|\mathbf{k}_f)
(\,{\sf I}-\mathbf{\hat k}_f\mathbf{\hat k}_f\,)
{\sf T}(\mathbf{ k}_f|\mathbf{ k}_i)\,\overline{\mathbf{ E}}_{\rm i},\\
\label{sex}
\sigma_{\rm e}&=&\frac{(2\pi)^3\mu_m}{k_m \overline{E}_{\rm i}^{2}}
\overline{\mathbf{ E}}_{\rm i}^{*}\cdot\frac{1}{2i}
\left[\, {\sf T}(\mathbf{ k}_i|\mathbf{ k}_i)
-{\sf T}^{\dagger}(\mathbf{ k}_i|\mathbf{ k}_i)\,\right]\,\overline{\mathbf{ E}}_{\rm i}.
\end{eqnarray}
\end{subequations}
Hence Eq.\ (\ref{im1}) implies the weaker conservation statement $\sigma_{\rm e}=\sigma_{\rm s}$ where the $T$-matrix is evaluated on-shell with $k_i=k_f=k_m$.

\section{Simplifications: partial summations of the Mie series}
\label{somaparti}
Part of the terms in the Mie series of the T-matrix can be explicitly summed. These are terms with no explicit frequency dependence. The calculation consists in identifying and reducing them, appealing to well-known sums that involve spherical Bessel functions to produce closed-form expressions.

\subsection{Longitudinal part}
\label{longproj}
Observe that the longitudinal part $T^{NN}$ involves the frequency-independent terms
\begin{eqnarray}
&&\sum_{l\geq 0}\left(\frac{k_1^2\varphi_{l,2}-k_2^2\varphi_{l,1}}{k_1^2-k_2^2}-1\right)\frac{j_l(a k_1)}{a k_1}\frac{j_l(a k_2)}{a k_2}\sum_{n=-l}^{l}\mathbf{ N}_{l n}(\Omega_{\mathbf{ k}_1})\mathbf{ N}_{l n}^*(\Omega_{\mathbf{ k}_2})\nonumber\\
\label{sumpar}
&=&\sum_{l\geq 0}\frac{k_1j_l(ak_1)j_l{\,'}(ak_2)-k_2j_l(ak_2)j_l{\,'}(ak_1)}
{a(k_1^2-k_2^2)}\sum_{n=-l}^l Y_{ln}(\Omega_{\mathbf{ k}_1})Y_{ln}^*(\Omega_{\mathbf{ k}_2})\mathbf{\hat k}_1\mathbf{\hat k}_2,
\end{eqnarray}
Since
\begin{eqnarray}
&&\sum_{ln}j_l(ak_1)j_l{\,'}(ak_2)Y_{ln}(\Omega_{\mathbf{ k}_1})Y_{ln}^*(\Omega_{\mathbf{ k}_2})=\frac{1}{(4\pi)^2}\frac{\partial}{\partial(ak_2)}\int {\rm d}\Omega_\mathbf{ x} e^{i a \mathbf{\hat x}.(\mathbf{ k}_1-\mathbf{ k}_2)}\nonumber\\
&=&\frac{1}{4\pi}\frac{\partial}{\partial(ak_2)}j_0(a|\mathbf{ k}_1-\mathbf{ k}_2|)
=-\frac{1}{4\pi}j_1(a|\mathbf{ k}_1-\mathbf{ k}_2|)\frac{k_2-k_1\mathbf{\hat k}_1.\mathbf{\hat k}_2}{|\mathbf{ k}_1-\mathbf{ k}_2|},
\end{eqnarray}
the sum in Eq.\ (\ref{sumpar}) evaluates to
\begin{equation}
\frac{1}{4\pi}\frac{j_1(a|\mathbf{k}_1-\mathbf{k}_2|)}{a|\mathbf{ k}_1-\mathbf{ k}_2|}(\mathbf{\hat k}_1.\mathbf{\hat k}_2) \mathbf{\hat k}_1 \mathbf{\hat k}_2.
\end{equation}
With $u=\mathbf{\hat k}_1.\mathbf{\hat k}_2$ and on account of Eq.\ (\ref{n1n2}), this longitudinal part reduces to
\begin{eqnarray}
\label{tpl}
&&{\sf T}^{NN}(\mathbf{ k}_1|\mathbf{ k}_2)\equiv\sum_{l\geq 0}
T_l^{NN}(k_1|k_2)\sum_{n=-l\dots l}\mathbf{ N}_{l n}(\Omega_{\mathbf{ k}_1})\mathbf{ N}_{l n}^*(\Omega_{\mathbf{ k}_2})\\
&=&\frac{3v}{(2\pi)^3}\frac{k_m^2}{\mu_m} \delta\varepsilon\Biggl[\sum_{l\geq 1} \frac{l(2l+1)\delta\varepsilon P_l(u)}{(\varepsilon_m/\varepsilon_s)\varphi_{l,s}- \varphi^{(1)}_{l,m}}\frac{j_l(a k_1)}{a k_1}\frac{j_l(a k_2)}
{a k_2}+ \frac{j_1(a|\mathbf{ k}_1-\mathbf{ k}_2|)}{a|\mathbf{ k}_1-\mathbf{ k}_2|}u\Biggr]\mathbf{\hat k}_1\mathbf{\hat k}_2,\nonumber
\end{eqnarray}
whence expression in Eq.\ (\ref{kegpar}) for $\mathbf{ k}_1=\mathbf{ k}_2=\mathbf{ k}$.

\subsection{Transverse part}
\label{pt}
For arbitrary $\mathbf{ k}_1$ and $\mathbf{ k}_2$, contributions involving VSHs $\mathbf{ Z}$ and $\mathbf{ X}$ do not simplify as easily. Still, for $\mathbf{ k}_1=\mathbf{ k}_2=\mathbf{ k}$, some partial evaluations of contributions to the Mie series are possible. Let us first write down a few useful limits, making use of the following derivatives:
\[\frac{\partial S_{l,ks}}{\partial k^2}=-\frac{1}{k^2-k_s^2}\left(S_{l,ks}-\frac{\partial\varphi_{l,k}}{\partial k^2}\right),\quad
\frac{\partial R_{l,ks}}{\partial k^2}=\frac{k_s^2}{k^2-k_s^2}\left(S_{l,ks}-\frac{\partial\varphi_{l,k}}{\partial k^2}\right).
\]
Thus,
\begin{subequations}
\label{limites}
\begin{eqnarray}
&&\lim_{k_1,k_2\to k}S_{l,12}=\frac{\partial\varphi_{l,k}}{\partial k^2},\\
&&\lim_{k_1,k_2\to k}R_{l,12}=-k^4\frac{\partial}{\partial k^2}\frac{\varphi_{l,k}}{k^2}
=\varphi_{l,k}-k^2\frac{\partial \varphi_{l,k}}{\partial k^2},\\
&&\lim_{k_1,k_2\to k}\frac{(k_1^2-k_m^2)S_{l,1s}-(k_2^2-k_m^2)S_{l,2s}}{k_1^2-k_2^2}\nonumber\\
&&\hspace{1cm}{}=\lim_{k_1,k_2\to k}\left[S_{l,12}+(k_s^2-k_m^2)\frac{S_{l,1s}-S_{l,2s}}{k_1^2-k_2^2}\right]
=\frac{\partial \varphi_{l,k}}{\partial k^2}+(k_s^2-k_m^2)\frac{\partial S_{l,ks}}{\partial k^2}\nonumber\\
&&\hspace{1cm}{}=S_{l,ks}-\frac{k^2-k_m^2}{k^2-k_s^2}\left(S_{l,ks}-\frac{\partial \varphi_{l,k}}{\partial k^2}\right),\\
&&\lim_{k_1,k_2\to k}\frac{k_2^2(k_1^2-k_m^2)R_{l,1s}-k_1^2(k_2^2-k_m^2)R_{l,2s}}{k_1^2-k_2^2}\nonumber\\
&&{}\hspace{1cm}{}=\lim_{k_1,k_2\to k}\left[k_s^2R_{l,12}+(k_s^2-k_m^2)\frac{k_2^2R_{l,1s}-k_1^2R_{l,2s}}{k_1^2-k_2^2}\right]\nonumber\\
&&\hspace{1cm}{}=-k_s^2 k^4\frac{\partial}{\partial k^2}\frac{\varphi_{l,k}}{k^2}
+(k_s^2-k_m^2)k^4\frac{\partial}{\partial k^2}\frac{R_{l,ks}}{k^2}\nonumber\\
&&\hspace{1cm}{}=k_m^2 R_{l,ks}+k^2k_s^2\frac{k^2-k_m^2}{k^2-k_s^2}\left(S_{l,ks}-\frac{\partial \varphi_{l,k}}{\partial k^2}\right).
\end{eqnarray}
\end{subequations}
The above limits allow us to write the transverse elements at equal momenta as:
\begin{subequations}
\label{kzeg}
\begin{eqnarray}
\label{eq:kzegzz}
&&T_l^{ZZ}(k|k)=
\frac{2a^3}{\pi}\frac{k_m^2}{\mu_m}
\Biggl\{\frac{(\delta\varepsilon R_{l,ks}+\delta\mu k^2 S_{l,ks})^2}
{(\varepsilon_m/\varepsilon_s)\varphi_{l,s}-\varphi_{l,m}^{(1)}}\\
&&\hspace{-2ex}{}+\frac{k^4}{k_m^2}\biggl[(k_s^2-k_m^2)\frac{\partial}{\partial k^2}\left(\delta\varepsilon\frac{R_{l,ks}}{k^2}+\delta\mu S_{l,ks}\right)+\delta\mu\frac{\partial\varphi_{l,k}}{\partial k^2}-\delta\varepsilon k_s^2\frac{\partial}{\partial k^2}\frac{\varphi_{l,k}}{k^2}\biggr]\Biggr\}\left[\frac{j_l(ak)}
{ak}\right]^2,\nonumber\\
\label{eq:kzegxx}
&&T_l^{XX}(k|k)=\frac{2a^3}{\pi}\frac{k^2}{\mu_m
}\Biggl[\frac{(\Delta\mu R_{l,ks}+k_m^2\Delta\varepsilon S_{l,ks})^2}
{(\mu_m/\mu_s)\varphi_{l,s}-\varphi^{(1)}_{l,m}}\\
&&{}-\frac{\mu_s}{\mu_m}(\Delta\mu k^2-k_m^2\Delta\varepsilon)^2\frac{\partial S_{l,ks}}{\partial k^2}+\Delta\mu k^4\frac{\partial}{\partial k^2}\frac{\varphi_{l,k}}{k^2}-k_m^2\Delta\varepsilon \frac{\partial \varphi_{l,k}}{\partial k^2}\Biggr]\left[\frac{j_l(ak)}{ak}\right]^2.\nonumber
\end{eqnarray}
\end{subequations}
The second and third terms of both these expressions depend on $l$ only via $\varphi_{l,k}$, $\partial \varphi_{l,k}/\partial k^2$ and $\varphi_{l,s}$. The last step consists in appealing to $\mathcal{S}(x)$ defined in Eq.\ (\ref{firstsum}), and to the following result:
\begin{equation}
\label{secondsum}
\sum_{l\geq 1}(2l+1)\frac{\partial \varphi_l(x)}{\partial x^2}[j_l(x)]^2=\frac{1}{3}\left[{\cal S}(x)-1\right]=-\frac{1}{15}x^2+O\left(x^4\right),
\end{equation}
whereas the evaluation of $\sum_{l\geq 1}(2l+1)\varphi_{l,s}[j_l(a k)]^2$ by means of elementary functions is most probably not feasible (this function should admit, for \emph{all} $l$, all the zeros of $j_l(ak_s)$ as poles relatively to the variable $k_s$). Using $\mathcal{S}(x)$ and Eq.\ (\ref{secondsum}) to sum up the terms of Eq.\ (\ref{kzeg}) that are independent of $\varphi_{l,s}$, one eventually arrives at Eqs.\ (\ref{keg}).

Finally, from $\mathcal{S}(x)$ and Eq.\ (\ref{secondsum}) one deduces the expansion
\begin{eqnarray}
\label{devlop}
&&\sum_{l\geq 1}(2l+1)\varphi_{l,s}\left[\frac{j_l(ak)}{ak}\right]^2=\frac{2}{3}{\cal S}(ak_s)
\nonumber\\
&&{}+\frac{1}{3k_s^2}\left[1-{\cal S}(ak_s)+2k_s^2\frac{\partial{\cal S}}{\partial k_s^2}(ak_s)\right](k^2-k_s^2)+O\left((k^2-k_s^2)^2\right),
\end{eqnarray}
which is useful to investigate the limit $k\to k_s$ alluded to in Sec.\ \ref{sec:eqmom}.



\begin{thebibliography}{10}
\markboth{Y.-P. Pellegrini, P. Thibaudeau and B. Stout}{Off-Shell T-matrix of dielectric and magnetic sphere}

\bibitem[1]{VARA80}
V.K. Varadan and V.V. Varadan, (eds.), {\em Acoustic, Electromagnetic and Elastic Wave Scattering}, Pergamon, Oxford, 1980.

\bibitem[2]{MISC04}
M.I. Mischenko, G. Videen, V. A. Babenko, N. G. Khlebtsov, and T. Wriedt, {\em T-matrix theory of electromagnetic scattering by particles and its applications: a comprehensive reference database}, J. Quant. Spectrosc. Radiat. Transf. 88 (2004), pp. 357--406.

\bibitem[3]{WATE07} P.C. Waterman, {\em The T-matrix revisited}, J. Opt. Soc. Am. A 24 (2007), pp. 2257--2267.

\bibitem[4]{MORO97} A. Moroz and A. Tip, {\em On-shell T-matrices in multiple scattering}, Phys. Lett. A 235 (1997), pp. 195--199.

\bibitem[5]{TSAN80}  L. Tsang and J.A. Kong, {\em Multiple scattering of electromagnetic waves by random distributions of discrete scatterers with coherent potential and quantum mechanical formulism}, {J. Appl. Phys.} {51} (1980), pp. 3465--3485.

\bibitem[6]{TSAN81}  L. Tsang and J.A. Kong, {\em Multiple scattering of acoustic waves by random distributions of discrrete scatterers with the use of quasi-crystalline coherent potential approximation}, {J. Appl. Phys.} {52} (1981), pp. 5448--5458.

\bibitem[7]{AGRA84} V.M. Agranovich and V.L. Ginzburg, {\em Crystal optics, spatial dispersion and excitons}, Springer-Verlag, Berlin, 1984.

\bibitem[8]{BARR07}  R.G. Barrera, A. Reyes-Coronado, and A. Garc\'{\i}a-Valenzuela, {\em Nonlocal nature of the electrodynamic response of colloidal systems}, Phys. Rev. B {75} (2007), 184202.

\bibitem[9]{SHEN95} Ping Sheng, {\em Introduction to wave scattering, localization and mesoscopic phenomena}, Academic Press, New York, 1995.

\bibitem[10]{FUCH92} R. Fuchs and P.\ Halevi, {\em Basic concepts and formalism of spatial dispersion}, in P.\ Halevi (ed.) {\em Spatial dispersion in solids \& plasmas}, Elsevier Science, New York, 1992, pp.\ 1--107.

\bibitem[11]{KARA64} F.C. Karal, Jr., and J.B. Keller, {\em Elastic, electromagnetic, and other waves in a random medium}, J. Math. Phys. 5 (1964), pp. 537--547.

\bibitem[12]{KELL66} J.B. Keller and F.C. Karal, Jr., {\em Effective dielectric constant, permeability, and conductivity of a random medium and the velocity and attenuation coefficient of coherent waves}, J. Math. Phys. 7 (1966), pp. 661--670.

\bibitem[13]{BERA70} M.J. Beran and J.J. McCoy, {\em Mean field variation in random media}, Quart. Appl. Math. 28 (1970), pp. 245--258.

\bibitem[14]{DIEN76} G. Diener and F. K\"aseberg, {\em Effective linear response in strongly heterogeneous media---self-consistent approach}, Int. J. Solids Structures 12 (1976), pp. 173--184.

\bibitem[15]{BART95} M. Barth\'el\'emy, H. Orland, and G. Z\'erah, {\em Propagation in random media: calculation of the effective dispersive permittivity by use of the replica method}, Phys. Rev. E 52 (1995), pp. 1123--1127.

\bibitem[16]{BALI95} R. Balian and J.-.J Niez, {\em Electromagnetic waves in random media: a supersymmetric approach}, J. Phys. I France 5 (1995), pp. 7--69.

\bibitem[17]{PELL97b} Y.-P.~Pellegrini,  D.B. Stout, and P. Thibaudeau, {\em Off-shell mean-field T-matrix of finite-size spheres and fuzzy scatterers}, {J.\ Phys.: Condens.\ Matter} {9} (1997), pp. 177--191.

\bibitem[18]{PELL97c} Y.-P.~Pellegrini, P. Thibaudeau, and D.B. Stout, {\em Wave propagation and spatial dispersion in random media}, in J.-F. Eloy (ed.) {\em Annales des Journ\'ees Maxwell 1995 (6--9 June 1995, Bordeaux-Lac, France)}, CEA-CESTA, Le Barp, 1996, pp. 335--338. {\texttt http://hal.archives-ouvertes.fr/hal-00412494/en/}.

\bibitem[19]{LIU90} J. Liu, L. Ye, D.A. Weitz, and Ping Sheng, {\em Novel acoustic excitations in suspensions of hard-sphere colloids}, Phys. Rev. Lett. 65 (1990), pp.\ 2602--2605.

\bibitem[20]{HESP01} L. Hespel, S. Mainguy, and J.-J. Greffet, {\em Theoretical and experimental investigation of the excitation in a dense distribution of particles: non-local effects}, J. Opt. Soc. Am. A 18 (2001), pp. 3072--3076.

\bibitem[21]{OBRI02}
S. O'Brien and J. Pendry, {\em Photonic band-gap effects and magnetic activity in dielectric composites}, J. Phys.: Condens Matter 14 (2002), pp. 4035--4044.

\bibitem[22]{PROS06}
S.L. Prosvirnin and S. Zhoudi, {\em On the effective constitutive parameters of metal dielectric arrays of complex-shaped particles}, J. of Electromagn. Waves and Appl. 20 (2006), pp. 583--598.

\bibitem[23]{KIRK85}  T.R. Kirkpatrick, {\em Localization of acoustic waves}, {Phys. Rev. B} {31} (1985), pp. 5746--5755.

\bibitem[24]{COND86}  C.A. Condat and T.R. Kirkpatrick, {\em Localization of acoustic waves}, {Phys. Rev. B} {33} (1986), pp. 3102--5755.

\bibitem[25]{LAND84} L.D. Landau, E.M. Lifshitz, and L.P. Pitaevskii, {\em Electrodynamics of continuous media}, Butterworth-Heinemann, Oxford, 1984.

\bibitem[26]{THIB97} P. Thibaudeau, {\em Contributions \`a la d\'etermination des propri\'et\'es \'electromagn\'etiques des milieux h\'et\'erog\`enes~: influence de la dispersion spatiale}, PhD.\ diss.\ (order number 1613), Uni\-ver\-si\-t\'e de Bordeaux I, France, 1997 (in French).

\bibitem[27]{PELL97a} Y.-P. Pellegrini, P. Thibaudeau, and D.B. Stout, {\em Momentum-dependent electromagnetic T-matrix and dynamic effective properties of random media}, {Physica A} {241} (1997), pp. 72--76.

\bibitem[28]{TIP97} A. Tip, {\em Nonconducting electromagnetic media with rotational invariance: transition operators and Green's functions}, J. Math. Phys. 38 (1997), pp. 3545--3570.

\bibitem[29]{PINH03} F.A. Pinheiro and B.A. van Tiggelen, {\em Light transport in chiral and magnetochiral media}, J. Opt. Soc. Am. A 20, pp. 99--105 (1997).

\bibitem[30]{TAI71}  Chen-To Tai, {\em Dyadic Green's Functions in Electromagnetic Theory}, Intext Educational Publishers, San Francisco, 1971.

\bibitem[31]{CHEW90}  W.C. Chew, {\em Waves and Fields in Inhomogeneous Media}, Van Nostrand Reinhold, New York, 1990.

\bibitem[32]{WATE71} P.C. Waterman, {\em Symmetry, unitarity, and geometry in electromagnetic scattering} Phys. Rev. D 3 (1971), pp. 825--839.

\bibitem[33]{JOAC75}  C.J. Joachain, {\em Quantum Collision Theory}, North-Holland, Amsterdam, 1975.

\bibitem[34]{FITZ95} R.M. Fitzgerald, A.A. Maradudin, and F. Pincemin, {\em Scattering of a scalar wave from a two-dimensional randomly rough Neumann surface}, Waves Random Complex Media 5 (1995), pp. 381--411.

\bibitem[35]{COHE87}  C. Cohen-Tannoudji, J. Dupont-Roc, and G. Grynberg, {\em In\-tro\-duc\-tion \`a l'\'e\-lec\-tro\-dy\-na\-mi\-que quan\-ti\-que}, Inter\'Editions/\'Editions du CNRS, Paris, 1987.

\bibitem[36]{JACK75}  J.D. Jackson, {\em Classical Electrodynamics, 2${}^{\rm nd}$ edition}, Wiley, New York, 1975.

\bibitem[37]{BOHR83}  C.F. Bohren and D.R. Huffman, {\em Absorption and Scattering of Light by Small Particles}, Wiley, New York, 1983.

\bibitem[38]{RUPP82} R.\ Ruppin, Chap.\ 9 in A.D.\ Boardman (ed.) {\em Electromagnetic surface modes} (Wiley, Chichester, 1982), pp. 345--398.

\bibitem[39]{NEWT66} R.G. Newton, {\em Scattering Theory of Waves and Particles}, McGraw-Hill, New York, 1966.

\bibitem[40]{OHTA80} K. Ohtaka, {\em Scattering theory of low-energy photon diffraction}, J. Phys. C 13 (1980), pp. 667--680.

\bibitem[41]{WATS80}  G.N. Watson, {\em A Treatise on the Theory of Bessel Functions}, Cambridge University Press, Cambridge, 1980.

\bibitem[42]{LAGE96} A. Lagendijk and B.A. Van Tiggelen, {\em Resonant multiple scattering of light}, Phys.\ Rep.\ 270 (1996), 143--215.

\bibitem[43]{KONG00} J.A. Kong, {\em Electromagnetic Wave Theory}, EMW Publishing, Cambridge, Mass. USA, 2000.

\bibitem[44]{WILL08} F. Willot, Y.-P. Pellegrini, M. Idiart, and P. Ponte Casta\~neda, {\em Effective-medium theory for infinite-contrast, two-dimensionally periodic linear composites with strongly anisotropic matrix behavior: dilute limit and crossover behavior}, Phys. Rev. B 78 (2008), 104111.

\markboth{Y.-P. Pellegrini, P. Thibaudeau and B. Stout}{Transition-matrix operator for momentum-dependent scattering by a spherical inclusion}
\end{thebibliography}
\end{document}